\documentclass[twocolumn]{aastex701}

\defcitealias{stefansson2025gaia}{S25}
\defcitealias{Barbato2026_gaps}{B26}
\defcitealias{Kanodia2026_gaia4}{K26}
\newcommand{\cta}{\citetalias}

\usepackage{graphicx}
\usepackage{subcaption}
\usepackage{CJK}
\usepackage{amsmath}
\usepackage[T1]{fontenc}

\begin{document}
\begin{CJK*}{UTF8}{gbsn}
\title{Epoch Astrometry and Radial Velocities Jointly Resolve the Resoeccentric Degeneracy in Gaia-4: \\ A Case Study and Forecast for \textit{Gaia} DR4}
\shorttitle{Resolving the Resoeccentric Degeneracy in Gaia-4 and in the \textit{Gaia} DR4 era}
\shortauthors{Yahalomi et al.}

\correspondingauthor{Daniel A. Yahalomi}
\email{yahalomi@mit.edu}

\author[0000-0003-4755-584X]{Daniel A. Yahalomi}
\altaffiliation{Juan Carlos Torres Postdoctoral Fellow}
\affiliation{Kavli Institute for Astrophysics and Space Research, Massachusetts Institute of Technology, Cambridge, MA 02139, USA}
\email{yahalomi@mit.edu}

\author[0000-0003-0834-8645]{Tiger Lu (陆均)}
\altaffiliation{Flatiron Research Fellow}
\affiliation{Center for Computational Astrophysics, Flatiron Institute, 162 Fifth Ave, New York, NY 10010, USA}
\email{tlu@flatironinstitute.org}

\author[0000-0001-6532-6755]{Quang H. Tran}
\altaffiliation{51 Pegasi b Fellow}
\affiliation{Department of Astronomy, Yale University, New Haven, CT 06511, USA}
\email{quang.tran@yale.edu}

\author[0000-0002-9544-0118]{Ben Cassese}
\affiliation{Minor Planet Center, Smithsonian Astrophysical Observatory, 60 Garden Street, Cambridge, MA 02138, USA}
\email{benjamin.cassese@cfa.harvard.edu}

\author[0000-0001-9907-7742]{Megan Bedell}
\affiliation{Center for Computational Astrophysics, Flatiron Institute, 162 Fifth Ave, New York, NY 10010, USA}
\email{mbedell@flatironinstitute.org}

\author[0000-0003-3130-2282]{Sarah C. Millholland} 
\affiliation{Kavli Institute for Astrophysics and Space Research, Massachusetts Institute of Technology, Cambridge, MA 02139, USA}
\affiliation{Department of Physics, Massachusetts Institute of Technology, Cambridge, MA 02139, USA}
\email{sarah.millholland@mit.edu}


\begin{abstract}
The stellar reflex motion induced by a single planet on an eccentric orbit is degenerate, to first order in eccentricity, with that of two planets on circular, coplanar orbits at a 2:1 period commensurability -- the \textit{resoeccentric degeneracy}. We fit models of a single eccentric planet and a near-2:1 pair to the RVs and the \textit{Gaia} DR4 prerelease epoch astrometry of Gaia-4, separately and jointly. The two architectures provide comparably good fits to the RVs alone and to the astrometry alone. However, a joint model of the astrometry and RVs definitively favors the single eccentric planet solution, implying a dynamically hot past for Gaia-4b. With a simple analytic argument, we estimate that roughly one fifth of the planets precisely characterized by \textit{Gaia} DR4 will be subject to the degeneracy. However, using this same analytic argument, we predict that the degeneracy can be resolved with sufficient RV follow-up for the majority of these systems. We provide a brief roadmap for identifying and resolving potentially degenerate systems with follow-up RVs, which will be necessary to accurately infer population-level statistics and evolutionary histories across the \textit{Gaia} DR4 catalog.
\end{abstract}



\section{Introduction}
\label{sec:introduction}

\textit{Gaia} Data Release 4 (DR4) arrives on 2 December 2026, and with it the first large sample of cool gas giants is projected to be discovered with time-series (epoch) astrometry for the first 5.5 yr of the \textit{Gaia} mission \citep{Gaia2016}. Injection-recovery forecasts predict $1{,}900\pm540$ planet detections with masses and periods characterized to better than 20\% precision \citep{LammersWinn_GaiaForecast}. At least an additional order of magnitude more stars are projected to show clear astrometric evidence of a nonluminous companion that cannot be precisely constrained with the data available in DR4 alone \citep{Yahalomi_epochalypse}.

To first order in eccentricity, the astrometric reflex signal induced by a single planet on an eccentric orbit is identical to that induced by two coplanar planets, of masses $M_{\rm in}$ and $M_{\rm out}$, on circular orbits with periods in a 2:1 ratio: the \textit{resoeccentric degeneracy} \citep{Yahalomi2026_resoeccentric}. The pair mimics a single planet with effective eccentricity 

\begin{equation}
    e_{\rm eff}\simeq2^{1/3}\,M_{\rm in}/M_{\rm out}.
    \label{eq:eeff}
\end{equation}
Mutual inclination between the two planets can theoretically break this degeneracy, although the specific inclination required depends on the system and precision of the astrometry \citep{Yahalomi2026_resoeccentric}. \citet{Greenfield_resoeccentric} recently showed with simulated \textit{Gaia} DR4 and DR5 astrometry that this degeneracy holds when $M_{\rm out} \gtrsim 3 M_{\rm in}$, corresponding to $e_{\rm eff} \lesssim 0.4$.

This degeneracy has been known in the radial velocity (RV) community for well over a decade \citep{Marcy2001_gj876, AngladaEscude2010}, and may affect a sizable fraction of known planets: for instance, \citet{Boisvert2018} argued a quarter of their sample of 60 single eccentric RV systems are in fact two-planet systems near a 2:1 resonance (at 95\% confidence). Injection-recovery tests show that the RV resoeccentric degeneracy appears to be most deceptive at moderate eccentricities, $e=0.32\pm0.12$ \citep[95\% interval 0.09--0.59;][]{Wittenmyer2019b}. This regime appears to be common among giant planets \citep[e.g.,][]{Blunt2026_ecc}. The first RV-confirmed substellar companions initially identified in \textit{Gaia} DR3 astrometry also follow this trend, each with $e\gtrsim0.26$ \citep{Holl2023, Sozzetti2023_gaia3b, stefansson2025gaia, Pinamonti2026_gaia6b, Barbato2026_gaps, Fitzmaurice2026_gaia3b, Kanodia2026_gaia4}.

While moderate eccentricities and 2:1 resonant orbits are both plausible outcomes of planet formation and subsequent dynamical sculpting, they are signatures of drastically different evolutionary pathways. Gas giants with moderate eccentricities likely form through dynamically hot channels \citep[e.g.,][]{RasioFord1996, juric2008dynamical, wu2011secular, weldon2025cold}. On the other hand, near-circular 2:1 pairs suggest more quiescent histories of convergent migration and resonant capture driven by planet-disk interactions \citep[e.g.,][]{LeePeale2002, lee2006resonance, Greenfield2026_resonance_stability} that subsequently remained long-term stable. Hence, the dynamical histories of planets in the DR4 catalog cannot be inferred from observations until the degeneracy is addressed system by system.

How, and with what data, then, can the resoeccentric degeneracy be broken? Gaia-4b is the optimal system on which this can be tested today. Gaia-4 is a late K or early M dwarf ($M_\star=0.644^{+0.025}_{-0.023}\,M_\odot$) at 73.7\,pc that was first identified as a candidate host of a substellar companion in \textit{Gaia} DR3, as one of the 72 \textit{Gaia} identified ``AStrometric Objects of Interest'' (ASOI) for being especially promising candidates for substellar companions \citep{Holl2023, GaiaCollaboration2023_Arenou}. \citet{stefansson2025gaia} presented RV observations for 28 of these candidates around M and K dwarfs, of which 21 turned out to be double-lined spectroscopic binaries, and confirmed Gaia-4b as a massive planet with $P=571.3^{+1.4}_{-1.3}$\,days, $M_{\rm p}=11.8\pm0.7\,M_{\rm Jup}$, and $e=0.338^{+0.026}_{-0.023}$ from a joint fit to their RVs and the DR3 orbital solution. Additional RVs were later presented in \citet{Barbato2026_gaps} and in \citet{Kanodia2026_gaia4}. 

Gaia-4 is also the only planet host among the three ``orbit'' sources in the DR4 prerelease of epoch astrometry and sits square in the resoeccentric degeneracy eccentricity danger zone as defined in \citet{Wittenmyer2019b}; the other two ``orbit'' sources in the prerelease data are the black-hole binary Gaia-BH3 \citep{GaiaBH3_discovery} and the low-mass stellar companion of HD~114762 \citep{Latham1989, Kiefer2019}. Thus, it provides a timely opportunity to combine astrometric and RV data in the planetary-mass regime and prepare a forecast of what is necessary to resolve the resoeccentric degeneracy in the \textit{Gaia} DR4 era.

In this Letter, we show that a joint fit of RVs and epoch astrometry can break the resoeccentric degeneracy. We fit a single eccentric planet and a near-2:1 planet pair to Gaia-4's RVs alone, to its DR4 prerelease epoch astrometry alone, and to both jointly. We perform a Bayesian model comparison of the nested-sampling evidences (Section~\ref{sec:data_models}). While the degeneracy is present both in the RVs and in the DR4 astrometry alone, it is resolved in favor of a single eccentric planet by the combination of the complete RV data set with the DR4 astrometry in the joint model (Section~\ref{sec:results}). We then provide intuition for why we were able to break the resoeccentric degeneracy in Gaia-4 and outline prospects of doing so for \textit{Gaia} DR4 identified planetary candidates (Section~\ref{sec:why}). Finally, in Section~\ref{sec:summary}, we translate the lessons learned into a roadmap for resolving the resoeccentric degeneracy in the DR4 era and summarize our findings.

\section{Data, models, and model comparison}
\label{sec:data_models}

\subsection{Radial Velocities}

We use the 23 published RVs of \citet[][hereafter \cta{stefansson2025gaia}]{stefansson2025gaia}, which include 10 from the Habitable-zone Planet Finder (HPF; \citealt{Mahadevan2012, Mahadevan2014}) spectrograph on the 9.2\,m Hobby-Eberly Telescope \citep{Ramsey1998, Hill2021}, 8 from the NEID \citep{Halverson2016, Schwab2016} spectrograph on the 3.5\,m WIYN telescope, and 5 from the FIber-fed Echelle Spectrograph (FIES; \citealt{Telting2014}) on the 2.56\,m Nordic Optical Telescope. We take these from their Table~5 as archived at VizieR (J/AJ/169/107).\footnote{\url{https://vizier.cds.unistra.fr/viz-bin/VizieR?-source=J/AJ/169/107}} They were obtained between 2023 April 5 and 2024 September 8, with median reported uncertainties of 15.9, 8.3, and 12.3\,m\,s$^{-1}$ for HPF, NEID, and FIES, respectively.\footnote{\cta{stefansson2025gaia} reported an eleventh HPF spectrum was obtained but excluded it for low signal-to-noise; we follow their published table.} 

We additionally include the 17 RVs presented in \citet[][hereafter \cta{Barbato2026_gaps}]{Barbato2026_gaps} from the High Accuracy Radial velocity Planet Searcher in the Northern hemisphere (HARPS-N; \citealt{Cosentino2012}) on the 3.6\,m Telescopio Nazionale Galileo. We take these from the CDS time series accompanying that paper, where Gaia-4 is listed as PM~J13580+3141. They span 2024 January 8 to 2025 July 7 and have a median reported uncertainty of 2.9\,m\,s$^{-1}$. The \cta{Barbato2026_gaps} velocities are absolute, so we subtract their median ($-17{,}431.3$\,m\,s$^{-1}$) to place them on the same scale as the relative \cta{stefansson2025gaia} velocities. \cta{Barbato2026_gaps}'s favored solution included a long-term quadratic trend.

Recently, \citet[][hereafter \cta{Kanodia2026_gaia4}]{Kanodia2026_gaia4} presented 4 additional RVs from HPF -- adding the low signal-to-noise observation excluded from the analysis of \cta{stefansson2025gaia} and 3 new observations on 2025 April 9, 2026 July 24, and 2026 July 29. \cta{Kanodia2026_gaia4} presented a joint model, fitting to all the RVs and the \textit{Gaia} prerelease astrometry (see Section~\ref{sec:gaia_prerelease}) simultaneously, spanning a $\sim$12-yr baseline, and found no significant evidence for the long-term quadratic trend found in \cta{Barbato2026_gaps}. In our analysis we do not include the new \cta{Kanodia2026_gaia4} HPF RVs, but rather use them as a test of the predictive power of our different model fits.

\subsection{\textit{Gaia} Epoch Astrometry} \label{sec:gaia_prerelease}

We use the \textit{Gaia} DR4 prerelease epoch astrometry\footnote{\url{https://www.cosmos.esa.int/web/gaia/dr4-prerelease}} of Gaia-4 ({\tt Gaia DR4 source\_id} 1457486023639239296). For each field-of-view transit (epoch) and each CCD, the file gives the along-scan (AL) centroid, its uncertainty, the scan angle, the AL parallax factor, and a flag ({\tt used\_by\_agis\_al}) recording whether the \textit{Gaia} Astrometric Global Iterative Solution (AGIS) used that measurement. We prepare these data in four steps:
\begin{enumerate}
\item We discard the sky-mapper measurement and keep the nine astrometric-field CCDs, leaving 968 measurements across 109 epochs. These sky-mapper measurements are for detection only and not precision astrometry \citep{Lindegren2012, Gaia2016, Lindegren2021}.
\item We remove every measurement that \textit{Gaia} flagged as unused ({\tt used\_by\_agis\_al = False}). This removes 144 measurements that were not used in the \textit{Gaia} AGIS, and 16 epochs drop out entirely. See \citet{Lindegren2012} for discussion of AGIS.
\item Within each remaining epoch, we remove CCD measurements that lie more than $5\sigma$ from the epoch median, where $\sigma$ is the larger of 1.4826 times the median absolute deviation of that epoch's CCDs \citep[the Gaussian-equivalent standard deviation;][]{Rousseeuw1993} and their median reported uncertainty, so that an epoch whose CCDs agree unusually well is not clipped against an artificially small scatter. This removes only one CCD measurement. A similar $5\sigma$ clipping per-CCD approach is used for outlier rejection in the \textit{Gaia} DR3 astrometric binary star pipeline \citep{Halbwachs2023}, and outliers are rejected prior to binning in \citet{Baycroft_kima}.
\item We combine the remaining 8--9 CCD measurements of each epoch into a single AL coordinate by inverse-variance weighting with the reported uncertainties. The observation time and scan angle are averaged with the same weights; the parallax factor is given once per epoch. A similar approach is adopted by previous injection-recovery studies \citep[e.g.,][]{LammersWinn_GaiaForecast, Yahalomi_epochalypse} and analyses of \textit{Gaia} prerelease data \citep[e.g.,][]{Baycroft_kima, Greenfield_resoeccentric}.

\end{enumerate}
We are left with 93 astrometric epochs between 2015 January and 2019 December, with a median uncertainty of 0.028\,mas per-epoch. We fit the one-dimensional AL coordinate directly, with times measured from J2016.0. The inverse-variance combination treats the CCDs within an epoch as independent.

\subsection{Combination of Different Datasets} \label{sec:combined_data}
\textit{Gaia} DR4 will come with thousands of astrometric planet candidates, likely followed by large-scale RV follow-up campaigns to confirm and characterize them. As such, it is beneficial to study what can be constrained with various subsets of the data. This provides us with a prediction for how characterization of potentially resoeccentric systems may take place in the DR4 era. As such, we test 5 different combinations of data:

\begin{enumerate}
    \item \textbf{\cta{stefansson2025gaia} RVs only:} the 23 FIES, HPF, and NEID RVs used in \cta{stefansson2025gaia}.
    
    \item \textbf{\textit{Gaia} DR4 prerelease astrometry only:} the 93 per-epoch astrometric observations.
    
    \item \textbf{\cta{stefansson2025gaia} + \cta{Barbato2026_gaps} RVs only:} the 17 new HARPS-N RVs presented in \cta{Barbato2026_gaps} and the 23 \cta{stefansson2025gaia} FIES, HPF, and NEID RVs.
    
    \item \textbf{\cta{stefansson2025gaia} RVs \& \textit{Gaia} DR4 prerelease astrometry jointly.}
    \item \textbf{\cta{stefansson2025gaia} + \cta{Barbato2026_gaps} RVs \& \textit{Gaia} DR4 prerelease astrometry jointly.}
    
\end{enumerate}

\subsection{Noise Model and Jitter}
\label{sec:noise_evidence}

We use independent Gaussian likelihoods for the RVs and astrometry and sum their log-likelihoods in the joint fits. Previous analyses (\cta{stefansson2025gaia}; \cta{Barbato2026_gaps}; \cta{Kanodia2026_gaia4}) found significant RV noise jitter in their models, with \cta{Kanodia2026_gaia4} finding jitter above the median uncertainty for all four instruments. Non-negligible jitter common to four independent spectrographs is unlikely to be instrumental, and suggests a physical origin -- likely either rotational modulation by spots (the photometric rotation period is 13.3\,days; \cta{stefansson2025gaia}) or additional companions. In order to understand the impact of including these fitted jitter terms, we perform fits both adopting the reported uncertainties (\textit{no jitter}) and with additional jitter terms for each RV instrument and astrometry added in quadrature (\textit{free jitter}). The \textit{free jitter} models can absorb a second planet signal and are hence biased towards the single planet solution, while the \textit{no jitter} models reward additional orbital parameters and favor the two-planet fit. Comparing both prescriptions tests the sensitivity of our conclusions to the noise model.

\subsection{Orbital Architectures and Evidences}

We compare two architectures: (1) a single planet on an eccentric orbit -- as modeled in \cta{stefansson2025gaia}, \cta{Barbato2026_gaps}, and \cta{Kanodia2026_gaia4} -- and (2) a near-2:1 pair modeled as two non-interacting Keplerian orbits, with a period ratio $P_{\rm in}/P_{\rm out} \sim \mathcal{N}(0.5, 0.1^2)$ truncated to $[0.05, 1]$.\footnote{The prior is intentionally much wider than that expected for formal 2:1 resonance. Therefore, a 2:1 period ratio is recovered only if the data favor it, as opposed to being confined there by a truncated prior. Thus, a posterior that settles far from the 2:1 resonance should be interpreted as additional evidence that the data do not favor a resoeccentric pair.} In the astrometric and joint models, both planets' inclination $i$ and longitude of ascending node $\Omega$ are independently fit. Both architectures are modeled with {\tt jaxoplanet} \citep{jaxoplanet} around a star of fixed mass $0.644\,M_\odot$ (\cta{stefansson2025gaia}). Following \cta{Kanodia2026_gaia4}, we propagate the uncertainty of the stellar mass into the planet masses after the fit: for each posterior sample we draw a stellar mass from a Gaussian centered on $0.644\,M_\odot$ with standard deviation $0.025\,M_\odot$, and solve the mass function again for the mass of each planet. The RVs and the astrometry constrain each planet's mass only through $M_{\rm p}/(M_\star+M_{\rm p})^{2/3}$, and thus this leaves the semi-amplitude $K$, the angular semi-major axis of the reflex orbit $a_0\equiv a_\star/d$, and the model comparison unchanged.

Although the resoeccentric degeneracy is exact to first order in eccentricity of the single planet only when assuming a circular 2:1 planet pair, we allow each planet in the 2:1 resonant solution small eccentricities. This is expected for a pair of planets captured into resonance via convergent migration, which pumps eccentricity mainly into the less massive planet -- in the resoeccentric degeneracy, this is the inner planet. At low eccentricities, the ratio of eccentricities for a pair of planets captured into the 2:1 mean motion resonance is approximated by Equation~7 of \citet{lee2004diversity}: $e_{\rm out}/e_{\rm in} \approx 0.29\,M_{\rm in}/M_{\rm out}.$ Plugging $e_\mathrm{eff} = 0.34$ (from the single-planet fit of \cta{stefansson2025gaia}) into Equation~\ref{eq:eeff} gives $M_{\rm in}/M_{\rm out}\approx0.27$ and thus yields the ratio $e_{\rm out}/e_{\rm in}\approx 0.08$. The specific value of $e_{\rm in}$ is set by the strength of eccentricity damping, with weaker damping corresponding to higher eccentricities \citep{deck2015migration}. For damping comparable to that required to reproduce the GJ~876 system (the archetypal migrated 2:1 pair with a similar mass ratio, with $e_{\rm in}\approx0.27$; \citealt{LeePeale2002, correia2010harps}), we expect $e_{\rm in}\lesssim0.3$ and adopt this as our bound. The ratio above then gives $e_{\rm out} \lesssim 0.025$ at lowest order. Higher-order terms may slightly raise this value, so we adopt $e_{\rm out}\lesssim 0.05$ for our prior. We have verified with \textit{N}-body \texttt{REBOUNDx} migration simulations\footnote{Using the prescription of \cite{tamayo2017convergent}.} \citep{Rein_2012, tamayo_reboundx} that these priors are sensible upper limits for Gaia-4.

All fits use the same four parameters for each orbit: its period, $\sqrt{e}\cos\omega$, $\sqrt{e}\sin\omega$ (where $\omega$ is the argument of periastron of the star's orbit), and the mean anomaly at the reference epoch. The period of the single (or outer) planet has a uniform prior between 474.46 and 674.46\,days. RV-only fits sample $K$ of each planet. Astrometry-only and joint fits instead sample the planet mass, $\cos i$, and $\Omega$, with independent orientations for the two planets of the pair, together with the five astrometric parameters of the star (two position offsets, two proper motions, and the parallax). In the joint fits, a single physical system therefore generates both the RVs and the AL coordinates. The joint RV and astrometric single eccentric planet and 2:1 planet pair fits with fitted jitter have 21 and 28 free parameters, respectively. All priors are listed in Table~\ref{tab:priors}.

We fit both architectures to five data combinations described in Section~\ref{sec:combined_data}. We sample each posterior and compute its evidence, $\mathcal{Z}$, with the importance nested sampler {\tt Nautilus} \citep{Lange2023_nautilus}. We use 1000 live points for both architectures in the fits to the \cta{stefansson2025gaia} RVs alone, 2000 for the single eccentric planet in the fit to both RV sets, and 2500 for the single eccentric planet in the astrometry-only and joint fits. The posterior of the 2:1 planet pair has more parameters and several modes, and we thus use 5000 live points for the remaining 2:1 architecture fits, except in the astrometry-only \textit{no jitter} fit and the joint fits with the \cta{stefansson2025gaia} RVs alone, where 2500 were enough for the runs to converge efficiently. Each run continues until the remaining live points hold less than 1\% of the evidence and the posterior reaches an effective sample size of $10^4$. All points drawn during the exploration phase are discarded.

We compare architectures using the Bayes factor $\ln\mathcal{B} = \ln\mathcal{Z}_{\tt two-planet } - \ln\mathcal{Z}_{\tt one-planet}$. In this convention, negative values favor the single eccentric planet. At an effective sample size of $10^4$, the statistical uncertainty of each $\ln\mathcal{Z}$ is approximately $N_{\rm eff}^{-1/2}\approx0.01$ \citep{Lange2023_nautilus}.

We additionally separate each Bayes factor into two terms. The first is the likelihood ratio of the two best-fitting solutions, $\Delta\ln\hat{\mathcal{L}} = \ln\hat{\mathcal{L}}_{\tt two-planet } - \ln\hat{\mathcal{L}}_{\tt one-planet}$, where $\hat{\mathcal{L}}$ is the highest likelihood among the posterior samples of a given fit. The second is the remainder, $\ln\mathcal{B} - \Delta\ln\hat{\mathcal{L}}$, which is the logarithm of the ratio of the Occam factors \citep[the fraction of each model's prior volume that is consistent with the data;][]{MacKay1992} and thus the penalty that the near-2:1 pair pays for the larger prior volume of its additional parameters.

We quote Bayes factors as $2\ln\mathcal{B}$ and likelihood ratios as $2\Delta\ln\hat{\mathcal{L}}$ throughout, such that the Bayes factors can be readily interpreted on the model comparison scale of \citet{KassRaftery1995} and the likelihood ratios are in the same units.

\begin{deluxetable}{ll}
\tablecaption{Priors for the single eccentric planet and 2:1 planet pair models. \label{tab:priors}}
\tablehead{\colhead{Parameter} & \colhead{Prior}}
\startdata
\multicolumn{2}{c}{All fits} \\
\hline
$P$ or $P_{\rm out}$ (days) & $\mathcal{U}(474.46,\,674.46)$ \\
$P_{\rm in}/P_{\rm out}$ & $\mathcal{N}(0.5,\,0.1^2)$ on $[0.05,\,1]$ \\
$e$ & $\mathcal{U}(0,\,0.95)$ \\
$e_{\rm out}$ & $\mathcal{U}(0,\,0.05)$ \\
$e_{\rm in}$ & $\mathcal{U}(0,\,0.3)$ \\
$\omega$ (rad) & $\mathcal{U}(0,\,2\pi)$ \\
$M_0$ (rad) & $\mathcal{U}(0,\,2\pi)$ \\
\hline
\multicolumn{2}{c}{RV-only fits} \\
\hline
$K$ (m\,s$^{-1}$) & $\log\mathcal{U}(0.1,\,5000)$ \\
\hline
\multicolumn{2}{c}{RV-only and joint fits} \\
\hline
$\gamma_j$ (m\,s$^{-1}$) & $\mathcal{U}(v_{j,\rm min}-\Delta_j,\,v_{j,\rm max}+\Delta_j)$ \\
$\sigma_{{\rm RV},j}$ (m\,s$^{-1}$)\tablenotemark{a} & $\log\mathcal{U}(0.5,\,50)$ \\
\hline
\multicolumn{2}{c}{Astrometry-only and joint fits} \\
\hline
$M_{\rm p}$ ($M_{\rm Jup}$) & $\log\mathcal{U}(0.1,\,80)$ \\
$\cos i$ & $\mathcal{U}(-1,\,1)$ \\
$\Omega$ (rad) & $\mathcal{U}(0,\,2\pi)$ \\
$\Delta\alpha^*$, $\Delta\delta$ (mas) & $\mathcal{U}(-500,\,500)$ \\
$\mu_{\alpha^*}$, $\mu_\delta$ (mas\,yr$^{-1}$) & $\mathcal{U}(-300,\,300)$ \\
$\varpi$ (mas) & $\mathcal{U}(1,\,30)$ \\
$\sigma_{\rm AL}$ (mas)\tablenotemark{a} & $\log\mathcal{U}(0.001,\,1)$ \\
\enddata
\tablenotetext{a}{\textit{Free jitter} models only.}
\tablecomments{$\mathcal{U}$ is uniform, $\log\mathcal{U}$ is log-uniform, and $\mathcal{N}$ is normal, truncated to the stated range. Subscripts ``in'' and ``out'' denote the planets of the 2:1 pair, with all other orbital parameters sampled for each planet. We sample $\sqrt{e}\cos\omega$ and $\sqrt{e}\sin\omega$ within the listed limits on $e$, and $M_0$ is the mean anomaly at J2016.0. For each RV instrument $j$, the offset prior spans its velocities, widened on each side by $\Delta_j = 0.5\,(v_{j,\rm max}-v_{j,\rm min}) + 100$\,m\,s$^{-1}$.}
\end{deluxetable}

\section{The resoeccentric degeneracy, and its resolution, in Gaia-4}
\label{sec:results}

In total, we fit 20 models to the available Gaia-4 observations, encompassing the two architectures, the five combinations of data, and the two noise treatments. We split the data in this way to learn what combination of observations break the degeneracy, and how resoeccentric candidates are best followed up in the \textit{Gaia} era.

Table~\ref{tab:solutions} presents all 20 fits. Figure~\ref{fig:degeneracy} shows the highest-likelihood solutions against the data with and without jitter and the comparison of Bayesian evidence of the two orbital architectures, Figure~\ref{fig:residuals} shows their residuals. Figures~\ref{fig:corner_1p}, \ref{fig:corner_2p}, and \ref{fig:corner_jitter} show the posteriors of the different models as corner plots.

\begin{deluxetable*}{llcccccccc}
\tablecolumns{10}
\tablewidth{0pt}
\tabletypesize{\footnotesize}
\tablecaption{Orbital solutions of the 20 fits: both architectures and five data combinations, each with free jitter and with no jitter. \label{tab:solutions}}
\tablehead{\colhead{Data} & \colhead{Noise} & \colhead{$P$} & \colhead{$K$} & \colhead{$M_{\rm p}$} & \colhead{$e$} & \colhead{$\omega$} & \colhead{$i$} & \colhead{$\Omega$} & \colhead{} \\ \colhead{} & \colhead{} & \colhead{(days)} & \colhead{(m\,s$^{-1}$)} & \colhead{($M_{\rm Jup}$)} & \colhead{} & \colhead{(deg)} & \colhead{(deg)} & \colhead{(deg)} & \colhead{}}
\startdata
\cutinhead{Single eccentric planet ({\tt one-planet})}
\cta{stefansson2025gaia} RVs & free jitter & $514^{+20}_{-16}$ & $354^{+16}_{-13}$ & \nodata & $0.262 \pm 0.035$ & $182.0 \pm 6.6$ & \nodata & \nodata &  \\
 & no jitter & $517.9^{+9.9}_{-8.8}$ & $338.0 \pm 8.0$ & \nodata & $0.251 \pm 0.016$ & $188.9 \pm 5.2$ & \nodata & \nodata &  \\
\noalign{\vskip 1.5pt\hrule height 0.25pt\vskip 1.5pt}
\textit{Gaia} DR4 astrometry & free jitter & $578.6 \pm 3.7$ & $332^{+31}_{-25}$ & $10.93 \pm 0.54$ & $0.404 \pm 0.078$ & $34.2 \pm 9.3$\tablenotemark{a} & $120.6 \pm 2.8$ & $174.5 \pm 2.7$\tablenotemark{a} &  \\
 & no jitter & $578.7 \pm 3.0$ & $332^{+24}_{-20}$ & $10.93 \pm 0.46$ & $0.402 \pm 0.062$ & $33.7 \pm 7.6$\tablenotemark{a} & $120.6 \pm 2.2$ & $174.3 \pm 2.1$\tablenotemark{a} &  \\
\noalign{\vskip 1.5pt\hrule height 0.25pt\vskip 1.5pt}
\cta{stefansson2025gaia} + \cta{Barbato2026_gaps} RVs & free jitter & $573.2^{+3.7}_{-3.3}$ & $318.2 \pm 3.3$ & \nodata & $0.327 \pm 0.014$ & $199.3 \pm 3.6$ & \nodata & \nodata &  \\
 & no jitter & $574.7 \pm 1.0$ & $317.67 \pm 0.80$ & \nodata & $0.3227 \pm 0.0042$ & $199.9 \pm 1.1$ & \nodata & \nodata &  \\
\noalign{\vskip 1.5pt\hrule height 0.25pt\vskip 1.5pt}
\cta{stefansson2025gaia} RVs \& & free jitter & $575.8 \pm 1.2$ & $321^{+12}_{-11}$ & $10.85 \pm 0.41$ & $0.328 \pm 0.022$ & $197.8^{+7.3}_{-5.9}$ & $120.2 \pm 2.2$ & $172.5 \pm 2.4$ &  \\
\textit{Gaia} DR4 astrometry & no jitter & $575.68 \pm 0.81$ & $321.2 \pm 5.8$ & $10.85 \pm 0.33$ & $0.323 \pm 0.011$ & $197.4 \pm 3.3$ & $120.1 \pm 1.5$ & $172.1 \pm 1.8$ &  \\
\noalign{\vskip 1.5pt\hrule height 0.25pt\vskip 1.5pt}
\cta{stefansson2025gaia} + \cta{Barbato2026_gaps} RVs \& & free jitter & $575.90 \pm 0.89$ & $317.3 \pm 3.3$ & $10.77 \pm 0.34$ & $0.321 \pm 0.011$ & $201.1 \pm 3.6$ & $120.5 \pm 1.7$ & $172.4 \pm 2.3$ &  \\
\textit{Gaia} DR4 astrometry & no jitter & $575.68 \pm 0.57$ & $317.66 \pm 0.79$ & $10.78 \pm 0.31$ & $0.3210 \pm 0.0039$ & $199.6 \pm 1.1$ & $120.6 \pm 1.3$ & $171.7 \pm 1.6$ &  \\
\cutinhead{Near-2:1 pair ({\tt two-planet})}
 &  & $P_{\rm out}$ & $K_{\rm out}$ & $M_{\rm out}$ & $e_{\rm out}$ & $P_{\rm in}/P_{\rm out}$ & $K_{\rm in}$ & $M_{\rm in}$ & $e_{\rm in}$ \\
 &  & (days) & (m\,s$^{-1}$) & ($M_{\rm Jup}$) &  &  & (m\,s$^{-1}$) & ($M_{\rm Jup}$) &  \\
\hline
\cta{stefansson2025gaia} RVs & free jitter & $502^{+18}_{-13}$ & $322.2 \pm 7.9$ & \nodata & $0.0349^{+0.0095}_{-0.0117}$ & $0.4923 \pm 0.0092$ & $67.4 \pm 9.1$ & \nodata & $0.167^{+0.077}_{-0.097}$ \\
 & no jitter & $515 \pm 19$ & $313.9^{+8.0}_{-6.9}$ & \nodata & $0.032^{+0.012}_{-0.014}$ & $0.4930 \pm 0.0066$ & $73 \pm 10$ & \nodata & $0.218^{+0.051}_{-0.067}$ \\
\noalign{\vskip 1.5pt\hrule height 0.25pt\vskip 1.5pt}
\textit{Gaia} DR4 astrometry & free jitter & $573.9^{+3.5}_{-3.9}$ & $286 \pm 15$ & $10.21 \pm 0.45$ & $0.029 \pm 0.015$ & $0.503 \pm 0.011$ & $57 \pm 33$ & $2.13 \pm 0.68$ & $0.136^{+0.102}_{-0.090}$ \\
 & no jitter & $575.2 \pm 2.6$ & $286 \pm 11$ & $10.19 \pm 0.38$ & $0.033^{+0.013}_{-0.016}$ & $0.5060 \pm 0.0084$ & $68 \pm 29$ & $2.32^{+0.56}_{-0.44}$ & $0.134 \pm 0.097$ \\
\noalign{\vskip 1.5pt\hrule height 0.25pt\vskip 1.5pt}
\cta{stefansson2025gaia} + \cta{Barbato2026_gaps} RVs & free jitter & $611.3 \pm 8.0$ & $306.4^{+8.7}_{-6.8}$ & \nodata & $0.030^{+0.013}_{-0.017}$ & $0.4622^{+0.0093}_{-0.0082}$ & $113.8 \pm 7.3$ & \nodata & $0.197 \pm 0.042$ \\
 & no jitter & $617.9 \pm 3.1$ & $315.8 \pm 2.9$ & \nodata & $0.0472^{+0.0021}_{-0.0043}$ & $0.4506 \pm 0.0030$ & $117.7^{+4.9}_{-4.1}$ & \nodata & $0.235 \pm 0.018$ \\
\noalign{\vskip 1.5pt\hrule height 0.25pt\vskip 1.5pt}
\cta{stefansson2025gaia} RVs \& & free jitter & $575.8 \pm 1.2$ & $300.4^{+7.5}_{-6.2}$ & $10.53 \pm 0.36$ & $0.032^{+0.013}_{-0.016}$ & $0.4944^{+0.0037}_{-0.0050}$ & $102.6 \pm 7.4$ & $2.86^{+0.33}_{-0.28}$ & $0.179 \pm 0.059$ \\
\textit{Gaia} DR4 astrometry & no jitter & $576.07 \pm 0.96$ & $299.5 \pm 4.6$ & $10.50 \pm 0.32$ & $0.030^{+0.014}_{-0.016}$ & $0.4952^{+0.0024}_{-0.0039}$ & $103.0 \pm 6.5$ & $2.89 \pm 0.26$ & $0.189^{+0.049}_{-0.055}$ \\
\noalign{\vskip 1.5pt\hrule height 0.25pt\vskip 1.5pt}
\cta{stefansson2025gaia} + \cta{Barbato2026_gaps} RVs \& & free jitter & $576.9 \pm 1.1$ & $293.8 \pm 5.0$ & $10.33 \pm 0.33$ & $0.031^{+0.013}_{-0.017}$ & $0.4973^{+0.0037}_{-0.0030}$ & $102.9 \pm 6.9$ & $2.85 \pm 0.30$ & $0.221 \pm 0.044$ \\
\textit{Gaia} DR4 astrometry & no jitter & $580.97 \pm 0.82$ & $303.0^{+1.9}_{-2.1}$ & $10.55 \pm 0.30$ & $0.0456^{+0.0034}_{-0.0089}$ & $0.4920 \pm 0.0020$ & $124.3^{+8.3}_{-16.4}$ & $3.48^{+0.36}_{-0.42}$ & $0.215^{+0.033}_{-0.025}$
\enddata
\tablecomments{Posterior medians with 16th--84th percentile intervals. The fits to RVs only sample $K$ and constrain no mass, inclination, or node. The fits that include the astrometry sample the planet mass at a fixed stellar mass of $0.644\,M_\odot$, and for them $K$ is derived from that mass, the period, eccentricity, and inclination. The masses listed include the uncertainty of the stellar mass, $0.644\pm0.025\,M_\odot$, propagated after the fit as in \cta{Kanodia2026_gaia4}. The prior confines the outer planet of the pair to $e_{\rm out}\le0.05$.}
\tablenotetext{a}{Astrometry alone determines $\omega$ and $\Omega$ only modulo $180^\circ$; the values are folded into $[0^\circ, 180^\circ)$.}
\end{deluxetable*}

\subsection{S25 Radial Velocities}
\label{sec:results_s25}

The 23 RVs used in \cta{stefansson2025gaia} span 522 days, less than one orbit of Gaia-4b. On their own, they do not distinguish the two architectures as the favored orbital architecture depends on the jitter assumption. With jitter, the evidence favors the single eccentric planet, $2\ln\mathcal{B}=-10.2$. Without jitter, it favors the pair, $2\ln\mathcal{B}=+3.7$. The reported uncertainties underestimate the scatter about either model: without jitter the best solutions have reduced $\chi^2$ of 3.5 and 3.3, and the fitted HPF and NEID jitters of 15--21\,m\,s$^{-1}$ are as large as the reported uncertainties. With the uncertainties fixed, only the more flexible model can absorb that excess scatter, and the 2:1 planet pair is rewarded for doing so.

\begin{figure*}
\centering
\includegraphics[width=\textwidth]{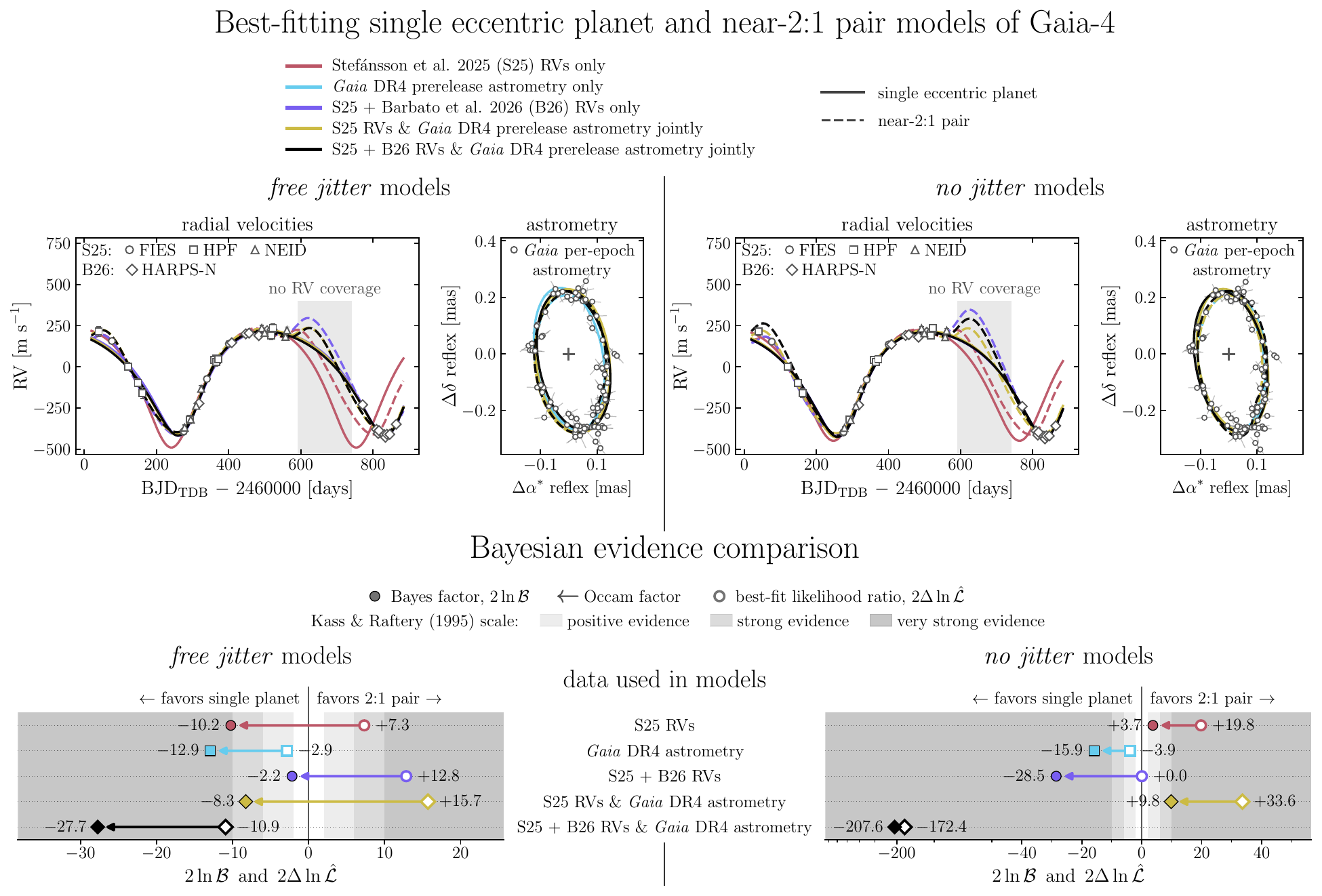}
\caption{The resoeccentric degeneracy in Gaia-4 for the \textit{free jitter} (left) and \textit{no jitter} (right) models. \textit{Top:} the RVs and the reflex orbit of the star on the sky, with the highest-likelihood posterior draw of each fit shown only in the panel whose data that fit used. \textit{Bottom:} the Bayesian evidence comparison for each data combination, with the evidence categories of \citet{KassRaftery1995} shaded. In the bottom right panel the axis is linear within $\pm60$ and logarithmic beyond.}
\label{fig:degeneracy}
\end{figure*}

The lower panels of Figure~\ref{fig:degeneracy} separate the Bayes factor into the likelihood ratio of the two best-fitting solutions and the remainder, which is the Occam factor that the pair pays for the larger prior volume of its additional parameters. For the \cta{stefansson2025gaia} RVs the pair fits better, by $2\Delta\ln\hat{\mathcal{L}}=+7.3$ with jitter and $+19.8$ without. The Occam factor is similar in the two cases, about $-17$ in $2\ln\mathcal{B}$. 

Our single planet fit to the \cta{stefansson2025gaia} velocities alone (Table~\ref{tab:solutions}) reproduces the RV-only solution of \cta{stefansson2025gaia} ($P=513^{+21}_{-16}$\,days, $e=0.263^{+0.037}_{-0.033}$, $K=355^{+17}_{-14}$\,m\,s$^{-1}$) to well within one standard deviation in every parameter.

\subsection{\textit{Gaia} DR4 Prerelease Astrometry}
\label{sec:results_astrometry}

The prerelease DR4 astrometry, similarly, does not strongly distinguish the architectures on fit quality. Both reproduce the 93 AL measurements with residual RMS of 0.032\,mas for each (Figure~\ref{fig:residuals}). However, the evidence favors the single eccentric planet, $2\ln\mathcal{B}=-12.9$ with jitter and $-15.9$ without. The best pair solution fits slightly worse than the best single-planet solution, by $2\Delta\ln\hat{\mathcal{L}}=-2.9$ with jitter and $-3.9$ without, and the remaining $-10.1$ and $-12.0$ in $2\ln\mathcal{B}$ are the Occam factor for the pair's seven additional parameters. 

\subsection{S25 \& B26 Radial Velocities}
\label{sec:results_harpsn}

The \cta{Barbato2026_gaps} velocities extend the RV baseline to 825 days and have reported uncertainties several times smaller. With all RVs from \cta{stefansson2025gaia} and \cta{Barbato2026_gaps}, the period agrees with the astrometric period, which the \cta{stefansson2025gaia} velocities alone placed three standard deviations too short.

The preference for the single planet weakens to $2\ln\mathcal{B}=-2.2$ including jitter, only showing ``positive evidence'' rather than ``very strong evidence'' in favor of the single planet model. The pair fits better than the single planet, by $2\Delta\ln\hat{\mathcal{L}}=+12.8$, which nearly cancels its Occam factor of $-15.0$. However, the optimal solution for the planetary pair settles far from resonance. Its outer period lengthens to $P_{\rm out}=611\pm8$\,days and its period ratio falls to $P_{\rm in}/P_{\rm out}=0.462^{+0.009}_{-0.008}$, four standard deviations from 2:1. Without jitter the two architectures fit equally well ($2\Delta\ln\hat{\mathcal{L}}=0.0$) and only including the Occam factor favors the single planet solution, with $2\ln\mathcal{B}=-28.5$. The gap in phase coverage of $\sim$210 days allows for flexibility in the near-2:1 pair solution, where a large disagreement between the single eccentric and 2:1 pair solution arises. A two planet model, fit to an intrinsic single eccentric system, will place the deviations in locations of significant data gaps in order to minimize disagreement between the observations and the model. Therefore it is likely in this case that if the data gap was in a different place, the deviations would be there instead. The RVs spanning more than a full orbit therefore constrain the pair tightly, but they do not exclude it, particularly when there are large data gaps. The outer period that the RVs require of the pair is, however, 3.7 standard deviations from the period that the astrometry measures independently for the single planet ($578.6\pm3.7$\,days), and 4.2 standard deviations from the one it measures for the outer planet of the pair ($573.9^{+3.5}_{-3.9}$\,days).

Our single planet fit to the \cta{stefansson2025gaia} and \cta{Barbato2026_gaps} velocities (Table~\ref{tab:solutions}) reproduces the Keplerian-only solution of \cta{Barbato2026_gaps} ($P=574.5^{+3.1}_{-2.8}$\,days, $e=0.336^{+0.013}_{-0.014}$, $K=310.0^{+2.7}_{-2.5}$\,m\,s$^{-1}$) to well within one standard deviation in $P$ and $e$, and to two standard deviations in $K$.

\subsection{S25 Radial Velocities \& \textit{Gaia} DR4 Prerelease Astrometry}
\label{sec:results_joint_s25}

A joint model of the \cta{stefansson2025gaia} RVs with the \textit{Gaia} DR4 astrometry is still somewhat ambiguous on the favored orbital architecture. The Bayes factor is $2\ln\mathcal{B}=-8.3$ with jitter and $+9.8$ without, the same dependence on the noise model as for the \cta{stefansson2025gaia} velocities alone. The pair fits better, by $2\Delta\ln\hat{\mathcal{L}}=+15.7$ with jitter and $+33.6$ without, against an Occam factor of about $-24$ in $2\ln\mathcal{B}$ in both cases.

\subsection{S25 \& B26 Radial Velocities \& \textit{Gaia} DR4 Prerelease Astrometry}
\label{sec:results_joint}

Adding the \cta{Barbato2026_gaps} RVs to the joint fit fully resolves the resoeccentric degeneracy. It gives $2\ln\mathcal{B}=-27.7$ with jitter and $-207.6$ without, very strong evidence for the single eccentric planet under either noise model. In the joint fit, the astrometry holds the outer period of the pair at $P_{\rm out}=576.9\pm1.1$\,days and its period ratio at $P_{\rm in}/P_{\rm out}=0.497\pm0.003$. When the outer planet is fixed at this period, the HARPS-N RVs strongly favor the single planet model. As a result, the 2:1 planet pair solution contains nearly twice as much jitter for the HARPS-N data as the single planet does ($21^{+5}_{-4}$ against $10.9^{+2.8}_{-2.1}$\,m\,s$^{-1}$). This is the only data combination in which the pair fits substantially worse than the single planet model ($2\Delta\ln\hat{\mathcal{L}}=-10.9$ with jitter and $-172.4$ without). The Occam factor contributes a further $-16.8$ and $-35.2$ in $2\ln\mathcal{B}$ in favor of the single planet model. We can thus state that the resoeccentric degeneracy in Gaia-4 is resolved in favor of a single eccentric planet solution, and we adopt for the discussion that follows the single eccentric planet from the joint fit to the astrometry and both RV sets, with jitter.

\subsection{Predicting the K26 Radial Velocities}
\begin{figure}
\centering
\includegraphics[width=\columnwidth]{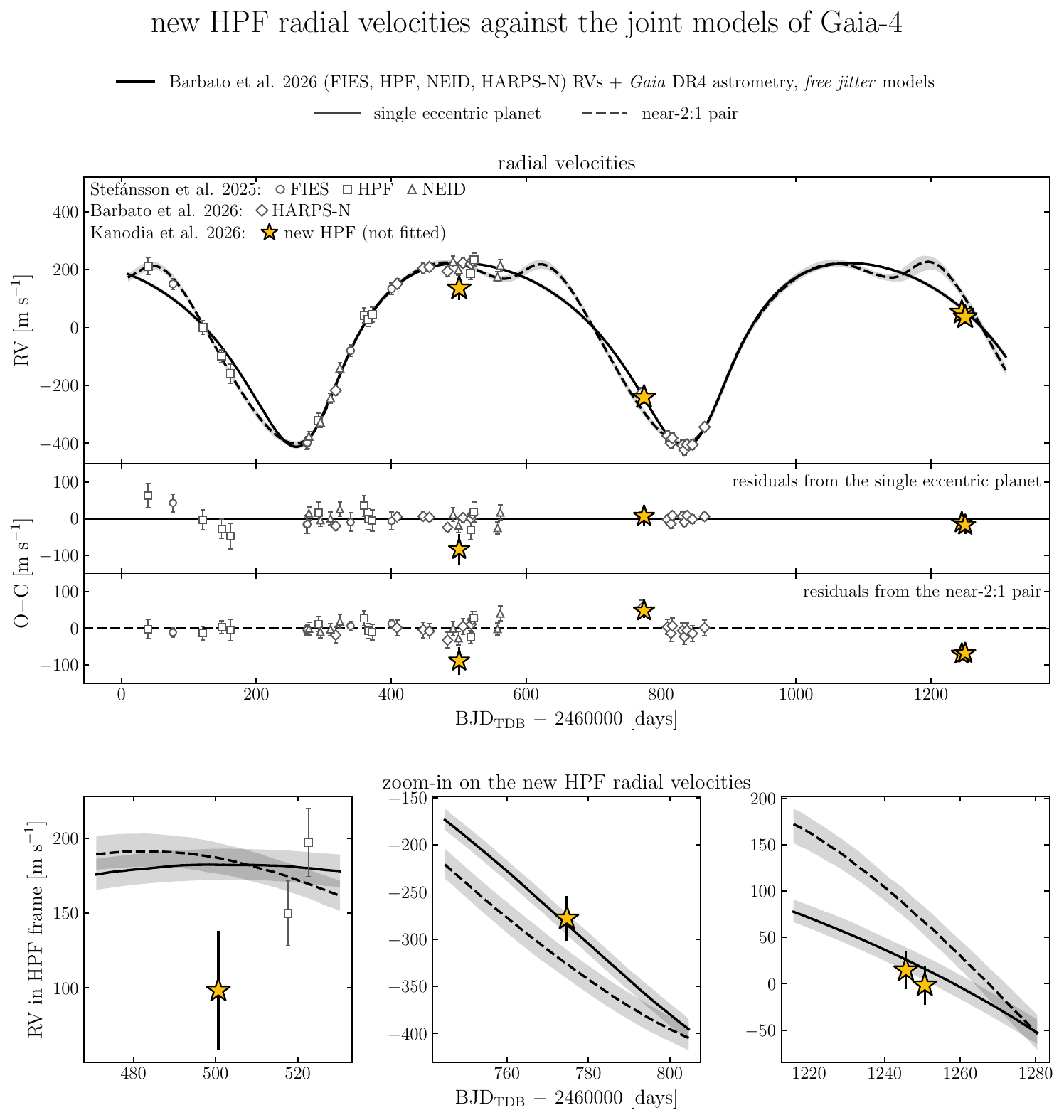}
\caption{Four HPF velocities published by \cta{Kanodia2026_gaia4} (yellow stars), which were not used in any of our fits, compared with the predictions of the single eccentric planet (solid) and the near-2:1 pair (dashed) from the \textit{free jitter} joint fits to the \cta{stefansson2025gaia} and \cta{Barbato2026_gaps} RVs and the astrometry. \textit{Top:} the posterior median predictions with 16--84\% bands and the RVs used in the fits (white markers). \textit{Middle:} residuals of the fitted (white markers) and new (bold stars) velocities. \textit{Bottom:} zooms on the four new velocities. Error bars include the fitted jitter.}
\label{fig:new_hpf}
\end{figure}

\cta{Kanodia2026_gaia4} recently published four additional HPF velocities that we did not use in any fit, two of them from 2026 July, a year after the last velocity in our fits. We take the \cta{Kanodia2026_gaia4} velocities from the repository accompanying that paper.\footnote{\url{https://github.com/shbhuk/pyGaiafits}} \cta{Kanodia2026_gaia4} rederived the HPF velocities, and thus the four new velocities are not on the zero point of the \cta{stefansson2025gaia} HPF velocities to which our fitted HPF offset refers. We measure the mean difference between the two reductions from the 10 HPF epochs in common, $+6.1\pm4.8$\,m\,s$^{-1}$ (\cta{Kanodia2026_gaia4} minus \cta{stefansson2025gaia}), subtract it from the four new velocities, and add its uncertainty in quadrature to their reported uncertainties. Figure~\ref{fig:new_hpf} compares them with the predictions of the joint fits. The single planet predicts them well and the pair does not. Thus, in addition to the very strong Bayesian model comparison evidence, the posterior-predictive ability of the single planet solution is much stronger than that of the 2:1 planet pair.

Our single planet joint fit to the \textit{Gaia} astrometry and the \cta{stefansson2025gaia} and \cta{Barbato2026_gaps} velocities (Table~\ref{tab:solutions}) reproduces the solution of \cta{Kanodia2026_gaia4} ($P=575.9\pm0.9$\,days, $e=0.319\pm0.012$, $K=315.1\pm3.8$\,m\,s$^{-1}$), which includes four additional HPF velocities and a linear and quadratic RV trend, to within one standard deviation in every parameter.

\section{Prospects for breaking the degeneracy in the Gaia DR4 era}
\label{sec:why}

The resoeccentric degeneracy is a mathematical happenstance that arises because the first order decomposition in eccentricity of the reflex motion caused by a single eccentric planet is equivalent to that of two circular, coplanar planets orbiting their host star with an orbital period ratio of 2:1. Taken from this perspective, it becomes clear that our capability to break the resoeccentric degeneracy lies in our ability to detect deviations from this first order decomposition. These deviations will primarily be caused by two physical realities:

\begin{enumerate}
    \item \textit{Single eccentric planets} will have second (and higher) order harmonics that will not be replicable by the 2:1 pair planets. Therefore, our ability to detect truly single planets depends on our ability to recover these higher order harmonic terms.

    \item \textit{2:1 resonant planets} will almost never sit at an exact 2:1 period ratio or be perfectly circular and coplanar, generating additional signals that a single eccentric planet cannot mimic. Therefore, our ability to detect truly 2:1 resonant planets depends on our ability to recover these deviations from circular, coplanar planets exactly in 2:1 resonance.
\end{enumerate}
Here we discuss both approaches and what each offers for \textit{Gaia} DR4 astrometrically identified planet candidates.

\subsection{Identifying Harmonic Structure of Truly Single Eccentric Planets}
\label{sec:guidance_scaling}

\subsubsection{Second-Order Harmonics and the Resoeccentric Corridor}
\label{sec:harmonic_theory}

\begin{figure*}
\centering
\includegraphics[width=\textwidth]{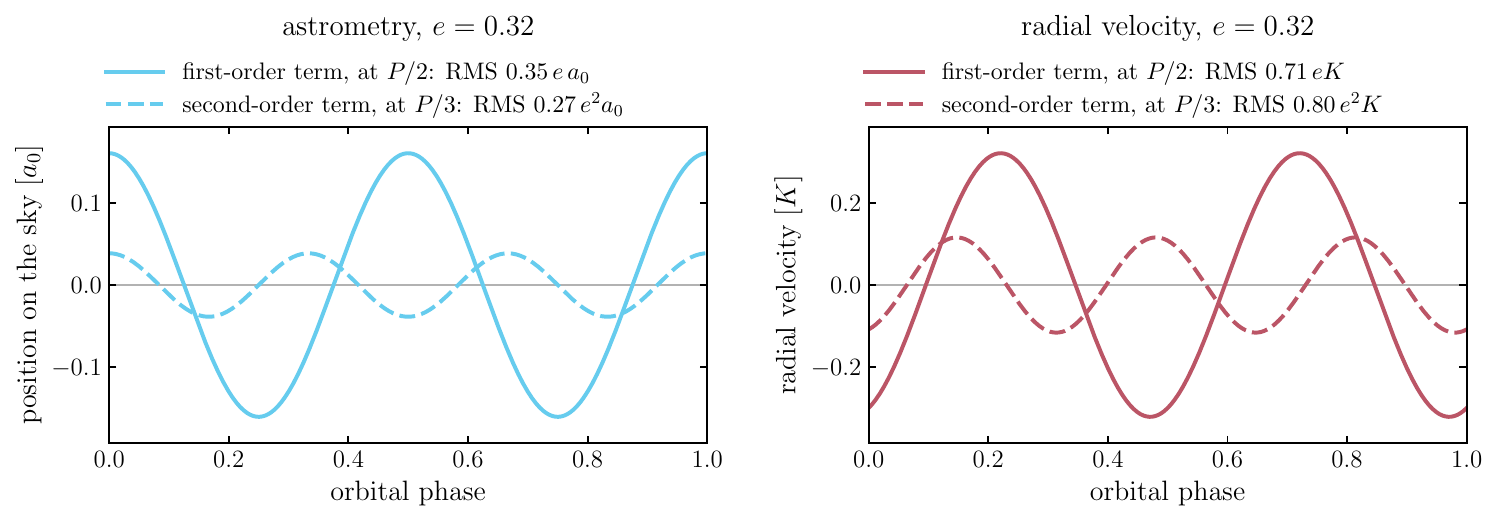}
\caption{The first-order (solid) and second-order (dashed) terms in $e$ of Equations~\ref{eq:x_expansion},  \ref{eq:y_expansion}, and \ref{eq:rv_expansion} for a single eccentric planet with the eccentricity of Gaia-4b ($e=0.32$), across a single orbit, in astrometry (left) and in RV (right). The solid curve is the term at $P/2$, which a single circular planet cannot replicate, and the dashed curve is the term at $P/3$, which a circular 2:1 pair cannot replicate. The astrometric signal is the position of the star along a single direction on the sky for a face-on orbit, in units of $a_0$, and the RV signal is in units of $K$. Our ability to constrain eccentricity depends on detecting the solid curve, and our ability to break the resoeccentric degeneracy for a single eccentric planet depends on detecting the dashed curve.}
\label{fig:harmonic_schematic}
\end{figure*}

We extend the derivation of the resoeccentric degeneracy in \citet{Yahalomi2026_resoeccentric} to second order in eccentricity. Expanding Kepler's equation as a power series to second order in eccentricity, the reflex motion of a star due to a single eccentric planet is
\begin{eqnarray}
\frac{x}{a_0}&=&\cos M-\tfrac32e+\tfrac12e\cos2M\nonumber\\
&&-\,e^2\left(\tfrac38\cos M-\tfrac38\cos3M\right),\label{eq:x_expansion}\\
\frac{y}{a_0}&=&\sin M+\tfrac12e\sin2M\nonumber\\
&&-\,e^2\left(\tfrac58\sin M-\tfrac38\sin3M\right),\label{eq:y_expansion}
\end{eqnarray}
where $x,y$ are the angular coordinates of the star in its orbital plane (with $x$ towards periastron), $M$ is the mean anomaly, and $\omega$ is the argument of periastron of the star's orbit. The physical displacement of the star along the line of sight is $z=a_\star\left[(x/a_0)\sin\omega+(y/a_0)\cos\omega\right]\sin i$, where $a_\star=a_0\,d$ is the semi-major axis of the star's orbit and $d$ is the distance to the star. The RV is the time derivative of $z$, which in terms of the semi-amplitude $K=2\pi a_\star\sin i/(P\sqrt{1-e^2})$ gives

\begin{eqnarray}
\frac{v}{K}&=&\cos(M+\omega)+e\cos(2M+\omega)\nonumber\\
&&+\,e^2\big[\tfrac98\cos(3M+\omega)-\cos(M+\omega)\nonumber\\
&&\quad-\,\tfrac18\cos(M-\omega)\big].\label{eq:rv_expansion}
\end{eqnarray}
Each eccentricity order in Equations~\ref{eq:x_expansion}, \ref{eq:y_expansion} and \ref{eq:rv_expansion} has a simple physical interpretation:

\begin{enumerate}
    \item The terms with no $e$ dependence describe a circular orbit at the orbital period, $P$.
    \item Apart from a constant offset, the first-order terms exactly describe the signal of a circular, coplanar inner orbit with period $P/2$, with a radius of $\tfrac12e\,a_0$ in astrometry and a semi-amplitude of $eK$ in RV. This is the resoeccentric degeneracy \citep{AngladaEscude2010, Yahalomi2026_resoeccentric}.
    \item The second-order terms slightly modify the amplitude and phase of the signal at $P$ and add a new harmonic at $P/3$, with a radius of $\tfrac38e^2a_0$ in astrometry and a semi-amplitude of $\tfrac98e^2K$ in RV.
\end{enumerate}
Because a planet on a circular orbit with a free mass, phase, and orientation can reproduce any sinusoidal motion of the star at its own orbital period, a circular 2:1 pair can absorb the signal of every term at $P$ and at $P/2$ but not at $P/3$. Our ability to break the resoeccentric degeneracy for a truly single eccentric planet comes down to detecting this second-order term. These terms at $P/2$ and $P/3$ can be seen in Figure~\ref{fig:harmonic_schematic}.

To compare the $P/2$ and $P/3$ terms with the precision of the data, we quantify the size of each by its root mean square (RMS) over one orbit: its typical strength in a single measurement, averaged over orbital phase. As the square of a sinusoid averages to one half over a cycle, the RMS of a sinusoid is its semi-amplitude divided by $\sqrt2$, and thus in RV the two terms have RMS amplitudes of $eK/\sqrt2\approx0.71\,eK$ and $\tfrac98e^2K/\sqrt2\approx0.80\,e^2K$.\footnote{See Appendix A of \citet{Lucy2005} for a similar derivation.} \textit{Gaia} measures the position of the star along a single scan direction at each epoch. For a face-on orbit, each term defines a circle of radius $r$ on the sky, and the position of the star on this circle has a component of $r\cos\phi$ along the scan direction, where $\phi$ is the angle between that position and the scan direction. The RMS of each term along scan is thus likewise $r/\sqrt2$, which gives $\tfrac12e\,a_0/\sqrt2\approx0.35\,e\,a_0$ and $\tfrac38e^2a_0/\sqrt2\approx0.27\,e^2a_0$. For an orbit of inclination $i$, each circle projects onto the sky as an ellipse with semi-axes of $r$ and $r\,|\cos i|$, and thus for uniformly distributed scan angles these RMS amplitudes are multiplied by an inclination factor, $f=[(1+\cos^2i)/2]^{1/2}$, which is 1 for a face-on orbit and $1/\sqrt2\approx0.71$ for an edge-on orbit.

We adopt a simple analytic criterion, in which a term of RMS amplitude $A$ is detected at $n\sigma$ confidence with $N$ measurements of uncertainty $\sigma$, evenly sampled in phase, if
\begin{equation}
A\geq n\,\sigma/\sqrt{N}.
\label{eq:detection}
\end{equation}
This criterion delineates three regimes for a given dataset. If the first-order term is not detected, then the signal is consistent with a circular orbit. If the second-order term is detected, then the data can distinguish a single eccentric planet from a circular 2:1 pair. In between, where the first-order term is detected and the second-order term is not, the interpretation of the system will be degenerate between a single eccentric planet and a 2:1 resonant pair, and we refer to this regime as the \textit{resoeccentric corridor}. As the corridor depends on the amplitude of the signal and the precision and number of the measurements, a planet can lie in the resoeccentric corridor in astrometry and outside of it in RV. We adopt a threshold of $3\sigma$ throughout the following analysis, since the eccentricity is a positive definite quantity, and thus noise alone biases the measured eccentricity of a circular orbit away from zero \citep{LucySweeney1971}. \citet{LammersWinn_GaiaForecast} find this bias in their mock \textit{Gaia} DR4 catalog: most of their planets with injected near-circular eccentricities return best-fits of $e\approx0.2$.

Setting the RMS amplitude of each term equal to the threshold of Equation~(\ref{eq:detection}) gives the $K$ and the $a_0$ at which the first-order ($K^{(1)}$ and $a_0^{(1)}$) and second-order ($K^{(2)}$ and $a_0^{(2)}$) terms are detected at $n\sigma$ significance,
\begin{eqnarray}
K^{(1)}=\frac{\sqrt2}{e}\,\frac{n\,\sigma}{\sqrt{N}},&\quad&
K^{(2)}=\frac{8\sqrt2}{9\,e^2}\,\frac{n\,\sigma}{\sqrt{N}},\label{eq:K_thresholds}\\
a_0^{(1)}=\frac{2\sqrt2}{e\,f}\,\frac{n\,\sigma}{\sqrt{N}},&\quad&
a_0^{(2)}=\frac{8\sqrt2}{3\,e^2f}\,\frac{n\,\sigma}{\sqrt{N}}.\label{eq:a0_thresholds}
\end{eqnarray}
The two RV thresholds cross at $e=8/9$, where the expansion to second order in $e$ breaks down.

\subsubsection{Analytic Predictions for \textit{Gaia} DR4 Planets}
\label{sec:harmonic_predictions}

For Gaia-4b, the first- and second-order terms have RMS amplitudes of 72 and 26\,m\,s$^{-1}$ in RV, and 23 and 5\,$\mu$as along scan, respectively. The reported uncertainties of its 40 RVs and 93 astrometric epochs are equivalent to 4.2\,m\,s$^{-1}$ per RV and 27\,$\mu$as per epoch, where we replace $\sigma/\sqrt{N}$ by $(\sum_i\sigma_i^{-2})^{-1/2}$ for unequal uncertainties. Equation~(\ref{eq:detection}) thus gives $3\sigma$ thresholds on the RMS amplitude of a term of 2.0\,m\,s$^{-1}$ and 8\,$\mu$as. Gaia-4b thus lies in the resoeccentric corridor in astrometry, where its first-order term is detected and its second-order term is not, whereas in RV both terms are detected and thus the degeneracy can be broken. However, in Section~\ref{sec:results_harpsn} we could not rule out the near-2:1 pair with the RVs alone. This is because our criterion is an analytic approximation with equally phase-spaced observations for a circular pair at exactly a 2:1 period ratio, whereas the near-2:1 pair that we fit in Section~\ref{sec:data_models} has a free period ratio and small eccentricities. 

\begin{figure*}
\centering
\includegraphics[width=\textwidth]{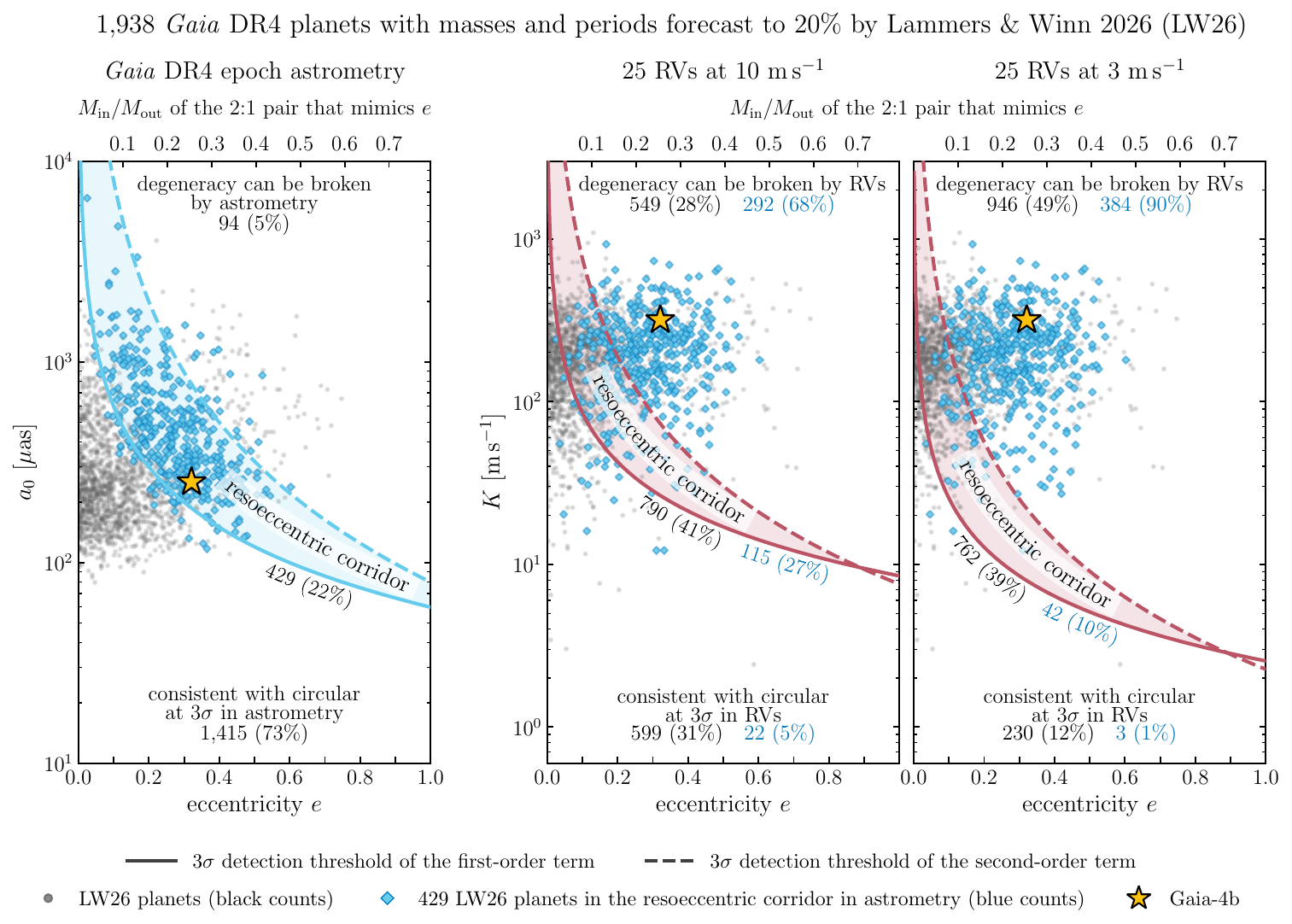}
\caption{The resoeccentric corridor for the 1,938 planets that \citet{LammersWinn_GaiaForecast} forecast to have masses and periods characterized to better than 20\% in \textit{Gaia} DR4. Analytic scalings for DR4 astrometry (left) and for 25 evenly phased RVs at 10 and 3\,m\,s$^{-1}$ (middle and right). The curves are $3\sigma$ detection thresholds of the first-order (solid) and second-order (dashed) terms in $e$ (Equations~\ref{eq:K_thresholds} and \ref{eq:a0_thresholds}). Blue diamonds and blue counts are the planets in the corridor in astrometry, and black counts are all planets. The astrometric curves assume the sample medians of $\sigma=54\,\mu$as and $N=88$ and the RMS of $f$ over isotropic orientations ($f=\sqrt{2/3}\approx0.82$), whereas each planet is classified with its own values, so 64 of the 429 blue diamonds fall outside the plotted corridor. The yellow star is Gaia-4b, which lies in the corridor in astrometry but not in RV.}
\label{fig:harmonic_regimes}
\end{figure*}

We apply this criterion to the 1,938 planets that \citet{LammersWinn_GaiaForecast} forecast to have masses and periods characterized to better than 20\% in \textit{Gaia} DR4 (Figure~\ref{fig:harmonic_regimes}), taken directly from their mock DR4 exoplanet catalog.\footnote{\url{https://github.com/CalebLammers/GaiaForecasts}} We compute $K$, $a_0$, and $f$ from the injected parameters of each planet. The solid (first-order) and dashed (second-order) curves of Figure~\ref{fig:harmonic_regimes} correspond to $n=3$ significance. The eccentricity of a planet below the solid curve is consistent with zero, a planet above the dashed curve is distinguished from a circular 2:1 pair, and a planet between the two lies in the resoeccentric corridor.

For the astrometry, we adopt the per-epoch uncertainty of \citet{LammersWinn_GaiaForecast}, which is 54\,$\mu$as for $G<14$ and increases for fainter stars. We take each host star's number of DR4 epochs, which their public catalog does not list, from the \texttt{epochalypse} catalog \citep{Yahalomi_epochalypse}, matched by \textit{Gaia} source ID. As in \citet{LammersWinn_GaiaForecast}, these epoch counts include the random rejection of 10\% of the predicted field-of-view transits, which accounts for occasional bad data points. 62/1938 of their host stars $({\sim}3\%)$ are not in the \texttt{epochalypse} catalog, as they are either fainter than \texttt{epochalypse}'s $G=16$ magnitude cut or lack a five-parameter \textit{Gaia} DR2 solution and thus are not in the \texttt{G23H} catalog \citep{Thompson_G23H} from which \texttt{epochalypse} stars are drawn. To these, we assign the median 88 epochs. For the RVs we consider a follow-up campaign of 25 velocities, evenly sampled in phase, with a precision of 10 or 3\,m\,s$^{-1}$. 

According to this criterion, with the DR4 astrometry eccentricity is not detected at $3\sigma$ for 1,415 (73\%) of the planets, which have a median eccentricity of 0.08. The astrometry alone can break the resoeccentric degeneracy for only 94 (5\%) of the planets, which have a median eccentricity of 0.47. The remaining 429 (22\%) lie in the resoeccentric corridor, with a median eccentricity of 0.27. Of the 429 astrometrically resoeccentric planets, we estimate that a follow-up campaign of 25 RVs at 10\,m\,s$^{-1}$ can break the degeneracy for 292 (68\%), and at 3\,m\,s$^{-1}$ for 384 (90\%). The systems that remain in the corridor in RV are those with lower eccentricities and smaller semi-amplitudes. It is thus evident that the resoeccentric corridor will be significant for DR4 detected planets and that adding RVs can break the degeneracy for most of these systems. This can be seen in Figure~\ref{fig:harmonic_regimes}.

This large fraction of planets without a detected eccentricity in \textit{Gaia} DR4 is consistent with \citet{LammersWinn_GaiaForecast}, who found that eccentricity will be constrained less precisely than the other orbital elements, with a typical relative error that is an order of magnitude larger than that on the planet mass. Taken from the perspective of the harmonic structure of the orbit, this is to be expected, as the first-order eccentricity term is only $e/2$ the size of the orbit. We can also compare our criterion to the orbit fits of \citet{LammersWinn_GaiaForecast} planet by planet, where we take half of the range between the 16th and 84th percentiles of each eccentricity posterior in their catalog as its $1\sigma$ uncertainty. We find that the injected eccentricity is more than $3\sigma$ away from zero, $e>3\sigma_e$, for 21\% of the planets in their fits, compared to 27\% with our criterion, and that the two agree on whether the eccentricity is detected for 92\% of the planets. Our criterion is thus slightly optimistic, likely due to the fact that it ignores correlations between the eccentricity and the other fitted parameters.

We emphasize that our conclusions are based on analytic estimates that calculate amplitudes only to leading order in eccentricity and assume complete phase coverage and uniformly distributed astrometric scan angles. Additionally, these analytic estimates implicitly assume that the alternative to a single eccentric planet is a circular planet pair at the exact 2:1 period ratio. Thus, they are upper limits on what such a campaign will deliver, suited for preliminary analysis that ranks targets for follow-up rather than detailed individual analysis.

\subsection{Identifying Non-Harmonic Structure of Truly 2:1 Resonant Planets}
\label{sec:guidance_baseline}

As previously discussed, planets in 2:1 mean-motion resonance generally are not perfectly circular. Additionally, planets in mean-motion resonance generally do not orbit at the exact period commensurability \citep{lithwick2012resonant, tamayo2025unified}, which causes the conjunctions to drift by $P_{\rm out}/P_{\rm in}-2$ inner cycles per outer orbit. Conjunctions of non-circular planets induce signals in the RV and astrometric data that are not reproduced by the harmonic decomposition of a single eccentric planet -- as can be seen in the ``conjunction bumps'' at $\sim$630 days and $\sim$1,200 days in the near-2:1 model for Gaia-4 RV data (see Figure~\ref{fig:degeneracy} and Figure~\ref{fig:new_hpf}). Observed at the correct location in phase space, these conjunction bumps can uniquely identify a 2:1 resonant system, and over a sufficient observing baseline this conjunction drift can and has been observed. For instance, \citet{Marcy1998_gj876} initially announced a single eccentric planet in the GJ~876 system, and after three more years of RVs were obtained, \citet{Marcy2001_gj876} reclassified the system as a 2:1 resonant pair and the drift in conjunctions can be clearly seen in their observations.

The baseline required to observe this drift is set by the fractional offset of the pair from the commensurability, $\Delta=P_{\rm out}/(2P_{\rm in})-1$, as the conjunctions drift through a full cycle in $P_{\rm out}/(2\Delta)$. For GJ~876, the apsidal lines, and with them the conjunctions, regress at roughly $-41^\circ$\,yr$^{-1}$ \citep{Millholland2018}, and a full cycle thus takes 8.8 yr, or roughly 53 orbits of the outer planet. DR4 will not have sufficient baseline to observe the conjunction drift en masse, as two thirds of the 1,938 forecasted planets considered above from \citet{LammersWinn_GaiaForecast} complete fewer than two orbits and nearly all complete fewer than five orbits over the 5.5 yr baseline. A pair of planets with conjunctions that drift at the same rate per orbit, with an outer period of 3.4 yr, the median of these \citet{LammersWinn_GaiaForecast} forecasted planets, would take 180 yr to complete a cycle. At the precision and baseline of \textit{Gaia} astrometry, recovering the inner resonant planet in GJ~876  will likely be a borderline case, and even so will likely be the outlier rather than the rule -- and detecting the conjunction drift for \textit{Gaia} DR4 2:1 resonant planets will hence likely take significant RV observations either from archival datasets in fortuitous situations or more commonly from targeted follow-up observations.

Planets in resonance also perturb each other's orbits, whereas throughout this work we model a near-2:1 pair as the sum of two non-interacting Keplerian orbits. This approximation is expected to hold for pairs near the 2:1 resonance \citep{Yahalomi2026_resoeccentric}, and \citet{Greenfield_resoeccentric} tested it directly by fitting non-interacting Keplerian orbits to simulated \textit{Gaia} astrometry of pairs evolved with $N$-body integrations. They found that the planet-planet interactions only become important for massive pairs on short orbital periods, of which \textit{Gaia} observes many orbits. As the vast majority of astrometric planets in DR4 will be observed for fewer than five orbits, the sum of two Keplerian orbits is thus likely an adequate model of a near-2:1 pair within the baseline of DR4.

\section{Summary}
\label{sec:summary}

\subsection{Joint Models and the Opposite Period Dependence of Astrometry and Radial Velocities}
\label{sec:guidance_period}

The amplitudes of a companion induced signal on its host star in RV and astrometric data scale in opposite directions with its orbital period. Explicitly, the RV semi-amplitude $K\propto M_{\rm p}P^{-1/3}$ and the astrometric semi-major axis $a_0\propto M_{\rm p}P^{2/3}$. Therefore, a 2:1 pair of mass ratio $q=M_{\rm in}/M_{\rm out}$ has a semi-amplitude ratio of $K_{\rm in}/K_{\rm out}=2^{1/3}\,q\approx1.26\,q$ but $a_{0,\rm in}/a_{0,\rm out}=2^{-2/3}\,q\approx0.63\,q$. Conversely, $a_{0,\rm out}/a_{0,\rm in}=2^{2/3}\,q^{-1}\approx1.59\,q^{-1}$ while $K_{\rm out}/K_{\rm in}=2^{-1/3}\,q^{-1}\approx0.79\,q^{-1}$ \citep[the former is Equation~18 of][]{Greenfield_resoeccentric}. Thus for 2:1 resonant planets, the inner planet is twice as prominent in RV as in astrometry while the outer planet is twice as prominent in astrometry as in RV.  Astrometry is thus dominant in constraining the outer orbit, while RVs dominate the characterization of the eccentric harmonic or inner planet, so a joint model benefits by using both datasets with opposing biases to constrain the same system. 

The astrometry fixes $P_{\rm out}$, $M_{\rm out}$, $i$, and $\Omega$, while the RVs decide whether the power at $P_{\rm out}/2$ is a harmonic or a second planet. This can be seen in the differing results for which models are favored for Gaia-4 when fit to different combinations of data. For the \cta{stefansson2025gaia} and \cta{Barbato2026_gaps} RVs alone, the 2:1 pair sat at $P_{\rm out}=611$\,days and $P_{\rm in}/P_{\rm out}=0.46$. In the joint model, the astrometry forced the outer period and period ratio to 577\,days and 0.497, respectively, where the model fails to accurately reproduce the harmonic structure of the eccentric single planet in the RVs (Section~\ref{sec:results_joint}).

\subsection{A Roadmap for \textit{Gaia} DR4}

For DR4 follow-up we recommend the following.
\begin{enumerate}
\item Use the resoeccentric corridor (Section~\ref{sec:guidance_scaling}) to flag single-planet solutions whose architecture may be degenerate with a 2:1 pair. We estimate that about 22\% of the planets forecast to be precisely characterized in DR4 \citep{LammersWinn_GaiaForecast} will be flagged.
\item Test each flagged system via model comparison between the two architectures using DR4 astrometry and all available archival RVs.
\item Follow up degenerate planets with RVs. We estimate that a campaign of 25 RVs at 10\,m\,s$^{-1}$ with complete phase coverage can break the degeneracy for 68\% of them, and at 3\,m\,s$^{-1}$ for 90\%.
\item Rank targets by the RV signal that a 2:1 pair cannot absorb, $0.80\,e^2K$, and prefer brighter hosts and shorter periods, where RVs are most sensitive.
\item Schedule the RVs prioritizing (1) complete phase coverage and (2) where the predictions of the two fitted architectures diverge most.
\item Fit both architectures to the RVs and the epoch astrometry jointly, with and without fitted jitter, and compare their evidences, as neither the RVs nor the astrometry of Gaia-4 could rule out a near-2:1 pair on their own.
\item If the 2:1 pair is preferred via model comparison, perform a dynamical analysis of the system and assess if the posteriors are consistent with stabilizing librating resonant angles.
\end{enumerate}

\subsection{Conclusion}

We fit a single eccentric planet and a near-2:1 pair to its RVs, to its \textit{Gaia} DR4 prerelease epoch astrometry, and to both jointly, with and without jitter. Via Bayesian model comparison, we confirm Gaia-4b's status as a single eccentric planet, rather than a pair of planets in 2:1 resonance. 

We have shown that the resoeccentric degeneracy in Gaia-4 is resolved when the RVs and \textit{Gaia} DR4 prerelease astrometry are fit jointly, even though it is present when each are fit separately. Neither the discovery RVs, the astrometry, nor the RVs extended with the more precise HARPS-N velocities can exclude a near-2:1 pair on their own. The joint fit to the astrometry and both RV sets, however, gives decisive evidence for the single eccentric planet under both noise models ($2\ln\mathcal{B}=-27.7$ in the \textit{free jitter} models and $-207.6$ in the \textit{no jitter} models), as the astrometry fixes the outer orbit, thus rejecting the longer outer companion period favored by the RV data alone in the 2 planet model. Deviation from the resoeccentric degeneracy enters at second order in eccentricity, $0.80\,e^2K$ in RV and $0.27\,fe^2a_0$ along scan in astrometry, which for Gaia-4b is a fifth of the uncertainty of one astrometric epoch. Thus, follow-up RV observations will be critical in resolving the resoeccentric degeneracy in \textit{Gaia} DR4 planet candidates.

Gaia-4 thus provides the first demonstration that the resoeccentric degeneracy can be resolved by combining RVs and epoch astrometry in a joint model. With the forthcoming \textit{Gaia} DR4, we expect thousands of cool giant planets with astrometric orbits \citep{LammersWinn_GaiaForecast}, and our expectations from RV demographic analysis of giant planets \citep{Blunt2026_ecc} and the eccentricities already recovered for the RV-confirmed DR3 candidates \citep{stefansson2025gaia, Barbato2026_gaps} suggest that many of these will sit in the moderate-eccentricity regime where a single eccentric planet most effectively mimics a 2:1 planet pair. 

Resolving the resoeccentric degeneracy clarifies Gaia-4b's dynamical history; given its eccentricity, the planet likely had a dynamically hot past. Although the eccentricity alone does not identify the specific excitation mechanism, detecting and characterizing additional companions could further constrain the system's evolution \citep[e.g.,][]{lu202614her}. Gaia-4b therefore represents an excellent case study of how resolving this architectural ambiguity enables a more robust inference of dynamical history. As the full DR4 sample is released, a catalog of seemingly single eccentric planets taken at face value will bias our planetary demographic statistics and our models of the formation and dynamical evolution of planetary systems. \textit{Gaia} DR4 will thus provide an abundance of targets for which RV follow-up, with full phase coverage and modeled jointly with the epoch astrometry, will be necessary to resolve the resoeccentric degeneracy system by system.

\software{
{\tt astropy} \citep{Astropy2013, Astropy2018, Astropy2022},
{\tt corner} \citep{ForemanMackey2016_corner},
{\tt JAX} \citep{jax2018github},
{\tt jaxoplanet} \citep{jaxoplanet},
{\tt matplotlib} \citep{Hunter:2007},
{\tt Nautilus} \citep{Lange2023_nautilus},
{\tt NumPy} \citep{numpy},
{\tt NumPyro} \citep{Bingham2019_pyro, Phan2019_numpyro},
{\tt pandas} \citep{McKinney2010_pandas},
{\tt REBOUND} \citep{Rein_2012},
{\tt REBOUNDx} \citep{tamayo_reboundx},
{\tt SciPy} \citep{Virtanen2020_scipy}
}

\begin{acknowledgments}
It is a pleasure to thank
Domenico Barbato,
Sarah Blunt,
Timothy Brandt,
Andrew Casey,
Jiayin Dong,
Daniel Fabrycky,
Emmanuel Greenfield,
Shubham Kanodia,
David Kipping,
Caleb Lammers, 
Malena Rice,
Ryan Rubenzahl,
Jessica Speedie,
Adrian Price-Whelan,
Gu{\dh}mundur Stef{\'a}nsson,
and Joshua Winn
for enlightening conversations. 

This work benefited from insightful conversations with the Astronomical Data Group at the Flatiron Institute, Center for Computational Astrophysics, the Millholland Research Group at MIT, and the Vanderburg Research group at Harvard.

D.A.Y. is supported by a Juan Carlos Torres Postdoctoral Fellowship at the Massachusetts Institute of Technology. T.L. is supported by a Flatiron Research Fellowship at the Flatiron Institute, a division of the Simons Foundation.

This work has made use of data from the European Space Agency (ESA) mission \textit{Gaia} (\url{https://www.cosmos.esa.int/gaia}), processed by the \textit{Gaia} Data Processing and Analysis Consortium (DPAC, \url{https://www.cosmos.esa.int/web/gaia/dpac/consortium}). Funding for the DPAC has been provided by national institutions, in particular the institutions participating in the \textit{Gaia} Multilateral Agreement. In particular, this work made use of the \textit{Gaia} DR4 prerelease astrometric time series (\url{https://www.cosmos.esa.int/web/gaia/dr4-prerelease}). We thank the \textit{Gaia} DPAC for making these resources available ahead of DR4.

The generative AI tool Claude (Fable and Opus) was used to assist in writing and debugging the code used for our model fits and data visualization, to check the manuscript for internal consistency and against the cited literature, and to assist in drafting and revising the manuscript text. All AI-assisted text, math, and code were reviewed, revised, and verified by the authors, who take full responsibility for the content of the manuscript.

\end{acknowledgments}



\bibliography{main}{}
\bibliographystyle{aasjournalv7}


\appendix

\section{Posteriors and Residuals}
\label{app:posteriors}

\subsection{Residuals}

\begin{figure*}[b]
\gridline{
\fig{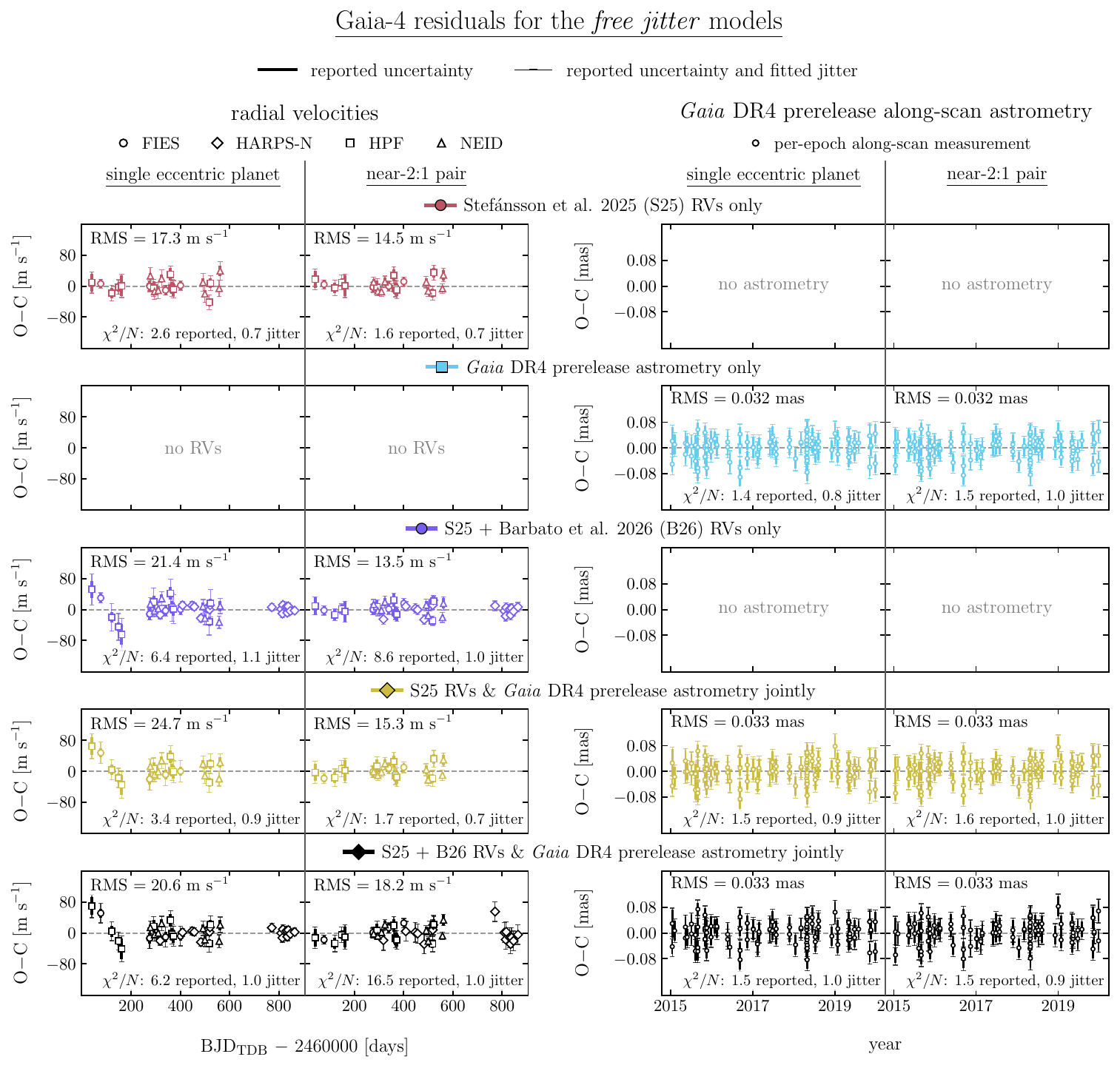}{0.45\textwidth}{(a) Free jitter}
\fig{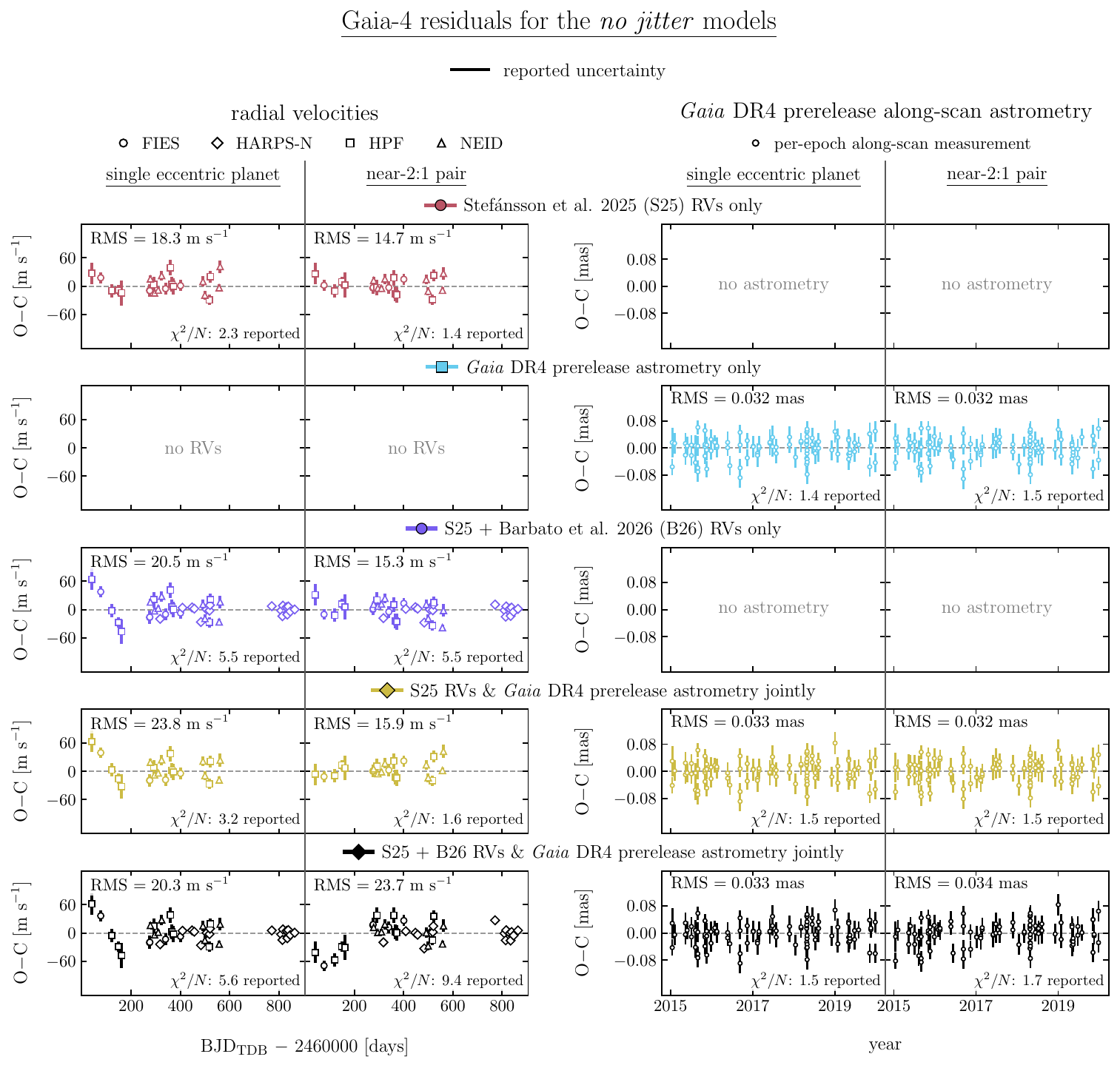}{0.45\textwidth}{(b) No jitter}}
\caption{Residuals of the solutions in Figure~\ref{fig:degeneracy} against the data each was fit to, (a) with fitted jitter terms and (b) without. Each row is one data combination. The left block shows the RV residuals and the right block the along-scan astrometric residuals; within each block the single eccentric planet is on the left and the near-2:1 pair on the right. Thick bars are the reported uncertainties. In (a), thin capped bars add that solution's fitted jitter in quadrature. Each panel gives the residual RMS and $\chi^2/N$ computed with the reported uncertainties and, in (a), with the jitter included.}
\label{fig:residuals}
\end{figure*}

Figure~\ref{fig:residuals} shows the residuals of the highest-likelihood solutions of Figure~\ref{fig:degeneracy} against the data each was fit to. The astrometric residuals do not distinguish the two architectures, with an along-scan RMS of 0.032--0.034\,mas in every fit, close to the median reported uncertainty of 0.028\,mas ($\chi^2/N=1.4$--$1.7$). The RV residuals have an RMS of 14--25\,m\,s$^{-1}$ and exceed the reported uncertainties for both architectures, with $\chi^2/N=1.4$--$3.4$ for the S25 velocities and $5.5$--$16.5$ once the more precise B26 velocities are included, and they reach $\chi^2/N\approx1$ only once the fitted jitter is added. The two architectures differ most in the joint fit to the astrometry and both RV sets, where the near-2:1 pair fits the HARPS-N velocities worse than the single eccentric planet, with a residual RMS of 19.8 against 10.5\,m\,s$^{-1}$ in the \textit{free jitter} models and 13.4 against 10.6\,m\,s$^{-1}$ in the \textit{no jitter} models. In the \textit{no jitter} models, the HARPS-N velocities account for $-112.5$ of the $2\Delta\ln\hat{\mathcal{L}}=-172.4$ between the two architectures, the other three RV instruments for $-42.7$, and the astrometry for $-17.2$.

\subsection{Corner Plots}

\begin{figure*}
\centering
\includegraphics[width=\textwidth,height=0.9\textheight,keepaspectratio]{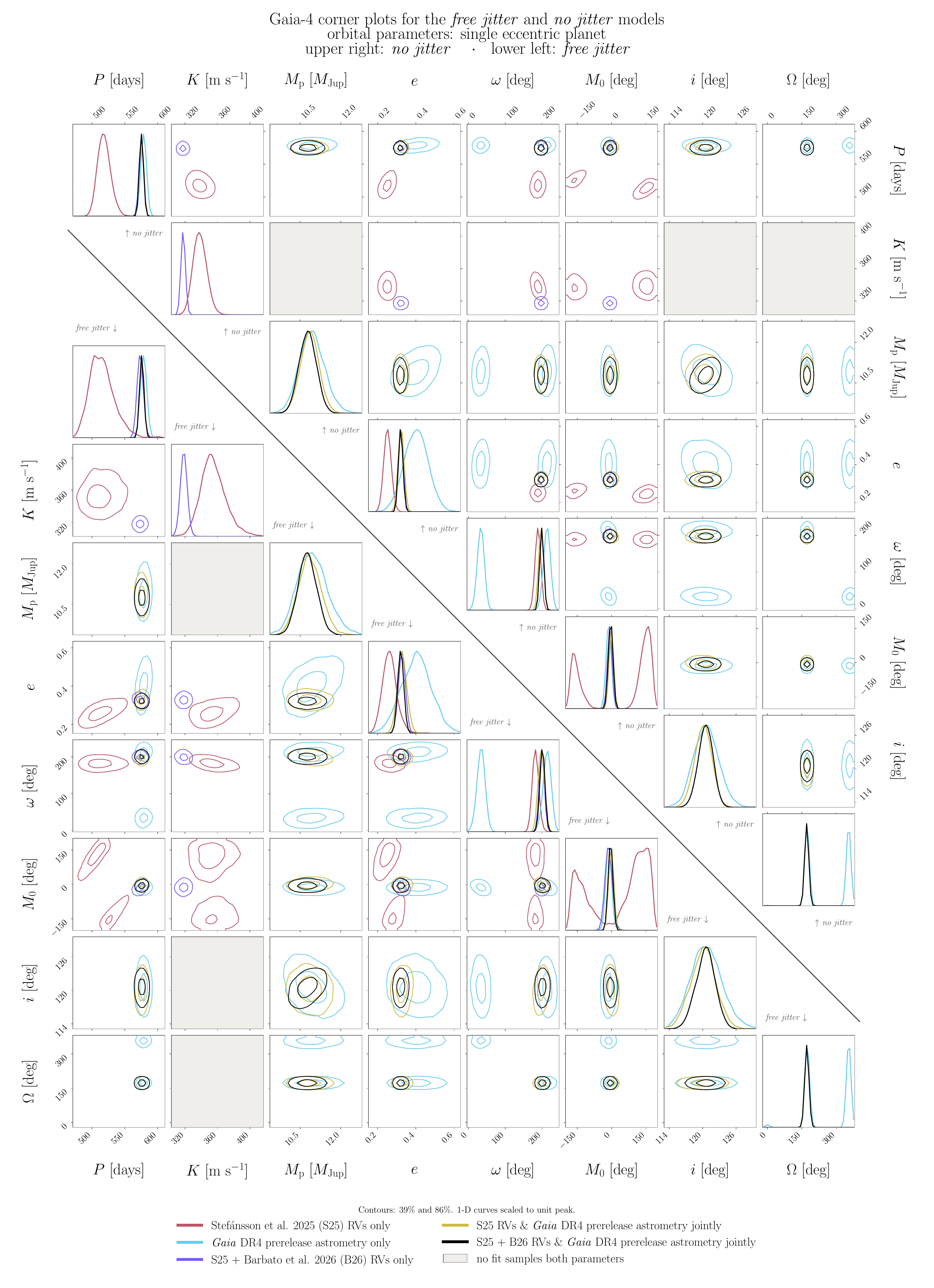}
\caption{Posterior distributions of the orbital parameters of the single eccentric planet for the five data combinations. The lower-left triangle shows the fits with free jitter and the upper-right triangle the fits without jitter. Contours enclose 39\% and 86\% of the posterior, and the one-dimensional distributions are scaled to unit peak. The semi-amplitude $K$ is sampled only in the RV-only fits, and the mass, inclination, and longitude of the ascending node only in the astrometry-only and joint fits; shaded panels pair parameters that no single fit samples.}
\label{fig:corner_1p}
\end{figure*}

\begin{figure*}
\centering
\includegraphics[width=\textwidth,height=0.9\textheight,keepaspectratio]{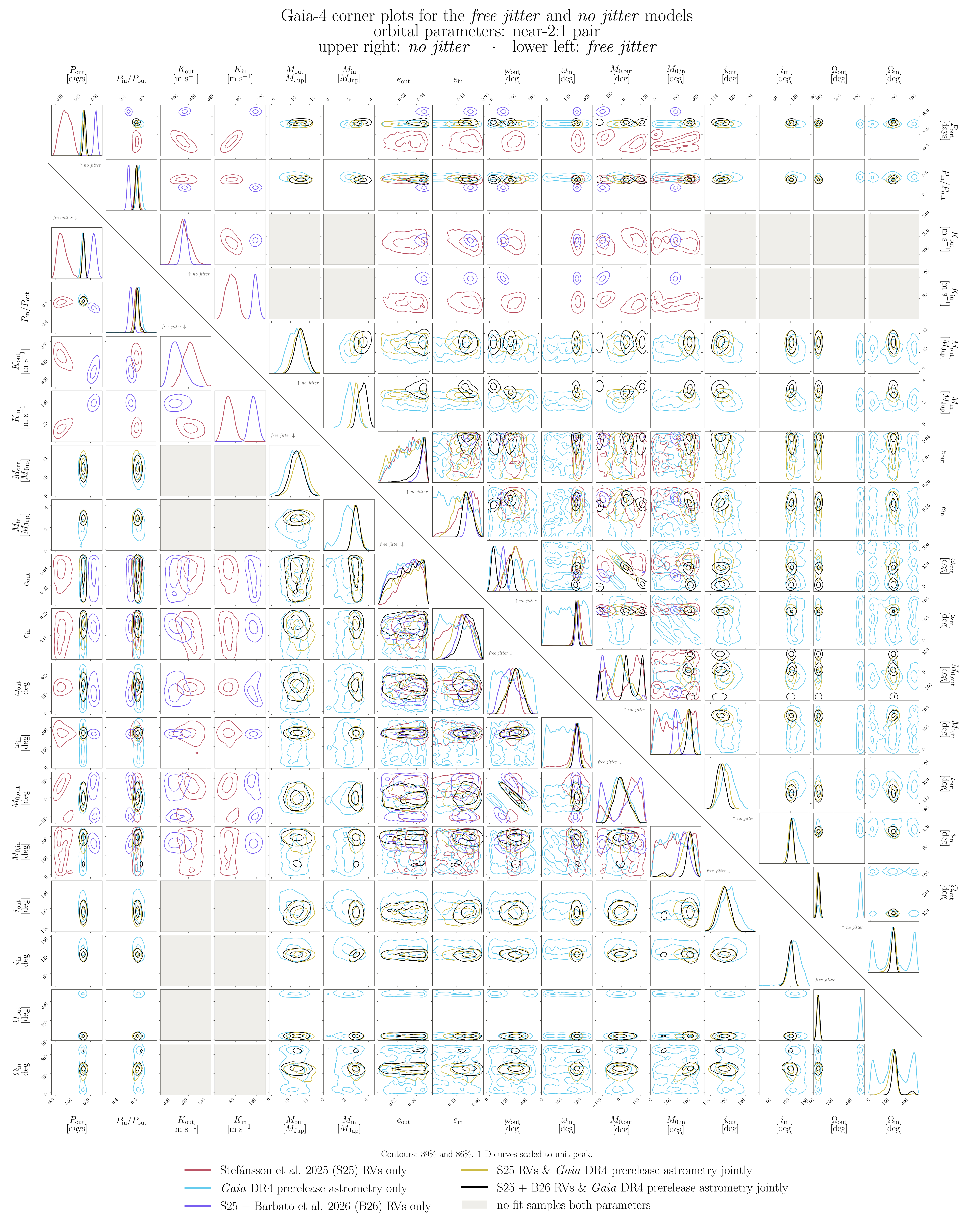}
\caption{Same as Figure~\ref{fig:corner_1p}, for the near-2:1 pair. Subscripts ``out'' and ``in'' denote the outer and inner planets.}
\label{fig:corner_2p}
\end{figure*}

\begin{figure*}
\gridline{
\fig{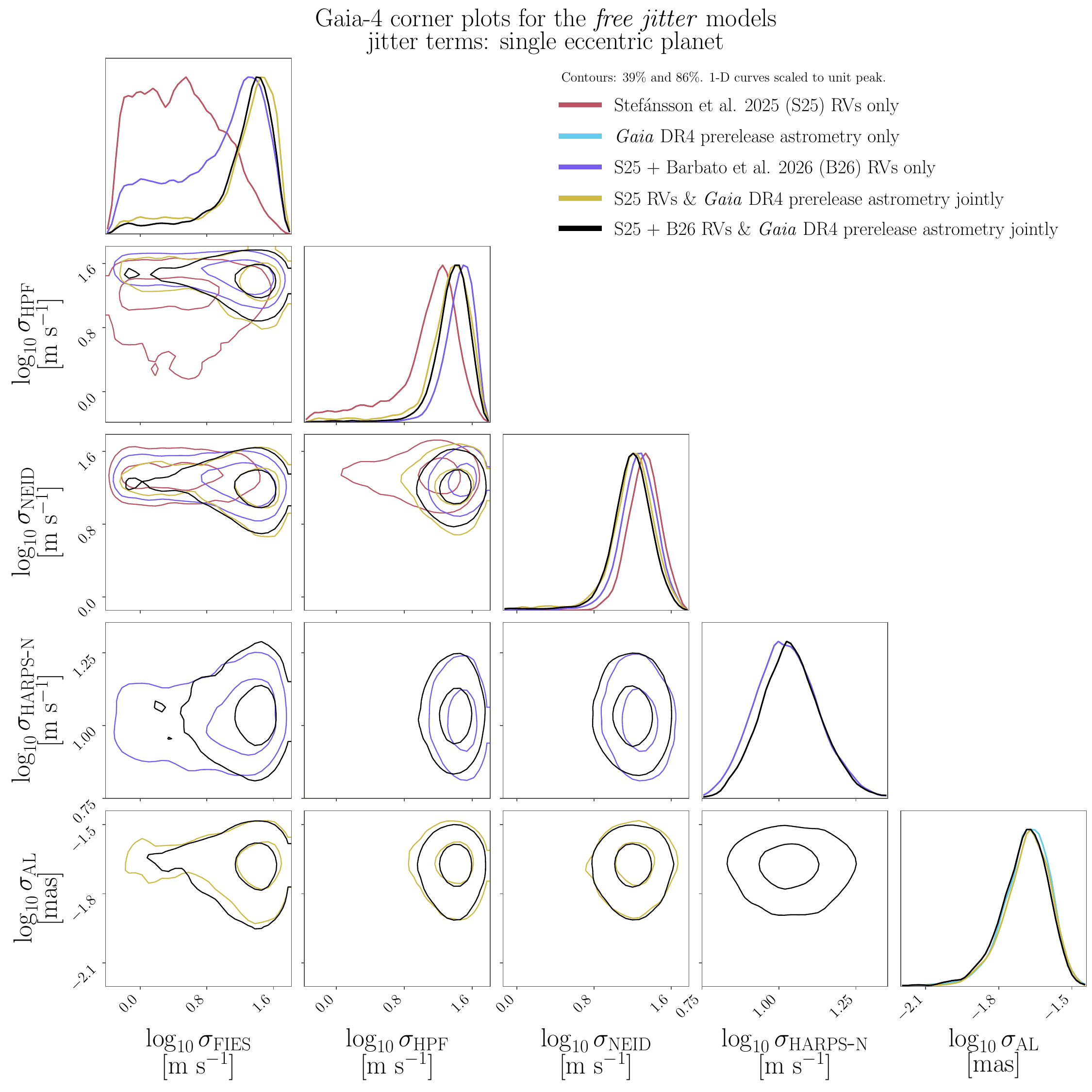}{0.49\textwidth}{(a) Single eccentric planet}
\fig{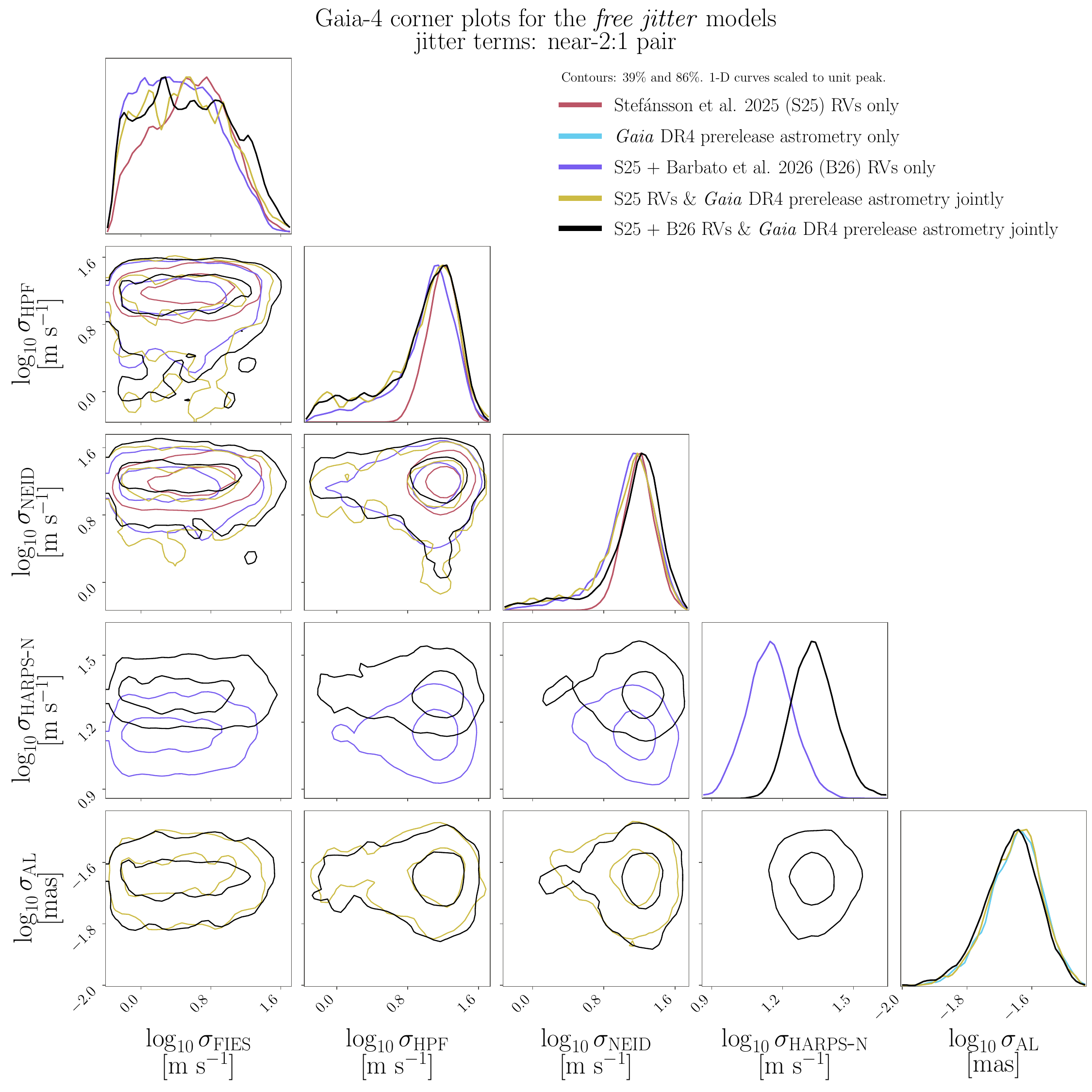}{0.49\textwidth}{(b) Near-2:1 pair}}
\caption{Posterior distributions of the jitter terms in the fits with free jitter, for (a) the single eccentric planet and (b) the near-2:1 pair. Each RV instrument has its own jitter term, and $\sigma_{\rm AL}$ is the along-scan astrometric jitter. Colors identify the data combination, and each fit appears only in the panels of the jitter terms it includes. Contours enclose 39\% and 86\% of the posterior, and the one-dimensional distributions are scaled to unit peak.}
\label{fig:corner_jitter}
\end{figure*}

Figures~\ref{fig:corner_1p} and \ref{fig:corner_2p} show the posteriors of the orbital parameters of the two architectures for the five data combinations, with and without jitter, and Figure~\ref{fig:corner_jitter} those of the jitter terms.

\end{CJK*}
\end{document}